\documentclass[apjl,iop,numberedappendix,twocolappendix]{emulateapj}

\usepackage{mathtools}
\usepackage{esint,graphicx,hyperref}
\hypersetup{colorlinks=false,pdfborder={0 0 0}}
\setcitestyle{notesep={; }}
\usepackage[normalem]{ulem}

\usepackage{graphicx}	
\usepackage{amsmath}	
\usepackage{amssymb}	
\usepackage{color}
\slugcomment{Accepted for publication in ApJ}

\newcommand{\iI}{{\mathcal{I}}}

\usepackage{lineno}

\begin{document}

\title{Spherical accretion in alternative theories of gravity}
\shorttitle{Spherical accretion in modified gravity}

\author{Adam~Bauer\altaffilmark{1,2} \and Alejandro~C\'ardenas-Avenda\~no\altaffilmark{1,2,3,4} \and Charles~F.~Gammie\altaffilmark{1,2,5,6} \and Nicol\'as~Yunes\altaffilmark{1,2}}
\shortauthors{Bauer et al.}
\altaffiltext{1}{Department of Physics, University of Illinois at Urbana-Champaign, 1110 West Green St, Urbana, IL 61801, USA}
\altaffiltext{2}{Illinois Center for Advanced Studies of the Universe, University of Illinois at Urbana-Champaign, 1110 West Green St, Urbana, IL 61801, USA}
\altaffiltext{3}{Programa de Matem\'atica, 
Fundaci\'on Universitaria Konrad Lorenz, 110231 Bogot\'a, Colombia.}
\altaffiltext{4}{Department of Physics, Princeton University, Princeton, NJ, 08544, USA}
\altaffiltext{5}{Department of Astronomy, University of Illinois at Urbana-Champaign, 1002 West Green Street, Urbana, IL 61801, USA}
\email{adammb4@illinois.edu} 
\altaffiltext{6}{National Center for Supercomputing Applications, 1205 W Clark St, Urbana, IL 61801, USA}


\begin{abstract}
The groundbreaking image of the black hole at the center of the M87 galaxy has raised questions at the intersection of observational astronomy and black hole physics. How well can the radius of a black hole shadow can be measured, and can this measurement be used to distinguish general relativity from other theories of gravity? We explore these questions using a simple spherical flow model in general relativity, scalar Gauss--Bonnet gravity, and the Rezzolla and Zhidenko parameterized metric. We assume an optically thin plasma with power-law emissivity in radius. Along the way we present a generalized Bondi flow as well as a piecewise-analytic model for the brightness profile of a cold inflow. We use the second moment of a synthetic image as a proxy for EHT observables and compute the ratio of the second moment to the radius of the black hole shadow. We show that corrections to this ratio from modifications to general relativity are subdominant compared to corrections to the critical impact parameter, and argue that this is generally true. In our simplified model the astrophysical parameter uncertainty dominates the gravity theory parameter uncertainty, underlining the importance of understanding the accretion model if EHT is to be used to successfully test theories of gravity.
\end{abstract}


\section{Introduction}

The Event Horizon Telescope (EHT) recently imaged the accretion flow around a supermassive black hole in the center of the M87 galaxy (hereafter, M87*) using very long baseline interferometry (VLBI) \citep{eht1, eht2, eht3, eht4, EHT5, eht6}. EHT images show a ``photon ring" and a ``black hole shadow" \citep{EHT5}. The photon ring corresponds to lines of sight with enhanced brightness due to their proximity to the black hole photon sphere. The black hole shadow refers to the central dark region of an image; this region corresponds to lines of sight that terminate at the black hole event horizon. An important result derived from EHT observational data was measuring the mass of M87*. The estimate assumed the validity of general relativity (GR) and a set of accretion models, and calibrated observational data from EHT to a library of simulated images.  The mass is inferred by comparing the observational and simulated data based on general relativistic magnetohydrodynamic (GRMHD) simulations \citep{eht6}. 
\par 
One science goal of EHT is to test GR against modified theories of gravity. EHT is capable of measuring the angular radius of the black hole shadow, denoted as $\theta_{s}$; for a non-rotating black hole in GR, $\theta_s := b_{c}/D = \sqrt{27} \ G M c^{-2} D^{-1}$, where $b_{c}$ is the critical impact parameter (boundary of the shadow), $D$ is the distance to the source, and $M$ is the black hole mass.  In M87*, $\theta_s \simeq 130 \ {\rm{picoradians}}$ \citep{eht6}. 

The shadow radius depends weakly both on spin and on the inclination of the spin vector relative to the line of sight. By measuring $\theta_s$ and the distance to the central black hole, one gains an estimate of the critical impact parameter $b_{c}$. The critical impact parameter is an intrinsic property of spacetime itself, i.e., it is independent of the astrophysical processes occurring within the spacetime. The critical impact parameter can be computed in non-GR theories, and one might therefore hope to use estimates of $b_{c}$ using EHT data to constrain departures from GR. The shadow radius is not directly observable, however, and must be inferred indirectly using a model for the observable millimeter wavelength emission, which typically peaks slightly outside $\theta_s$ \citep{eht6}. The EHT collaboration used numerical models and analytic estimates to constrain the impact that deviations from the Kerr metric could have on the size of the black hole shadow while remaining consistent with EHT observations \citep{EHTmod}. \cite{kocherlakota} takes a similar approach, using shadow size to infer constraints on physical charges (or hairs) in modified gravity theories. Beyond EHT's current capabilities, related analyses to test GR have been suggested by studying the photon ring structure at high angular resolution~\citep{Johnson:2019ljv,2020PhRvD.102l4004G}. 

The analysis conducted in \cite{EHTmod}, however, has been challenged \citep{gralla}. One critique is that the EHT is only capable of measuring the characteristic angular radius of the emitting plasma; from this, $\theta_s$ is inferred by comparing the observed characteristic radius of emission to realizations of $\theta_s$ in a suite of GRMHD-based simulated VLBI data products.  It is possible, however, that the measurement of the characteristic radius of emission is far more sensitive to differences in astrophysical model choices than GR modifications, making a deviation from GR difficult to detect amidst significant uncertainty in astrophysical model parameters \citep{2019PhRvD.100b4039C,sebas, glampedakis,2021arXiv211000026L}. A second critique is that \cite{EHTmod} assumed that the measurement of $M$ made with EHT (which relied on the validity of GR) holds even if the metric is modified. \cite{gralla} argued that one cannot assume GR and an accretion model to measure the mass, and then turn around and assume this mass and accretion model to test GR. Put plainly, \cite{gralla} advocates for a direct calibration of EHT data with modified GRMHD codes, which could then be used to test GR. 

In this paper we reconsider these critiques in simple spherical accretion models.  In each model we calculate the intensity of the image on the sky $I$ as a function of impact parameter $b$.  EHT's observing technique sparsely samples Fourier components of the image, and the ring size is inferred through a complicated procedure that we cannot fully consider here \citep{eht6}.  Instead, we will use the second angular moment of the image to define a characteristic radius of the source
\begin{equation}
r^{2}_{char} := \dfrac{2 \pi \int_{0}^{\infty} b^{3} I(b) db}{2\pi \int_{0}^{\infty} b I(b) db}. \label{eq:rchardef}
\end{equation}
This is equivalent to using only the information in the shortest baselines (lowest spatial frequency Fourier components) sampled by EHT.  The notion is that $r_{char}$ is an observable that can be used to compute $b_c$ by dividing by a dimensionless ratio 
\begin{equation}
\vartheta := \dfrac{r_{char}}{b_{c}}. \label{eq:varthetadefn}
\end{equation}
that is determined from an astrophysical model.  Again, we need to emphasize that this is not the procedure used by EHT, but as we shall see it is instructive nonetheless.   
We consider spherical inflow in general relativity, in scalar Gauss-Bonnet (sGB) gravity, and in a theory-agnostic ``bumpy'' black hole parameterization introduced by Rezzolla and Zhidenko (RZ) \citep{Rezzolla:2014mua}.  By computing images for each model we can understand how changes in the astrophysical model and the gravitational theory (or parameterized changes in the gravitational field) translate into changes in $\vartheta$ and therefore propagate into uncertainty in the inferred shadow radius.   

We will find that $\vartheta$ is far more sensitive to changes in the emission model than a deviation in the spacetime metric. Furthermore, it is insensitive to the velocity profile of the accreted plasma in Bondi accretion. The characteristic radius of emission and the critical impact parameter do, individually, depend on the metric parameters that represent deviations from the Schwarzschild spacetime. The \textit{ratio} $\vartheta$, however, is highly insensitive to these parameters. We will show that the location of the characteristic radius of emission and the critical impact parameter covary with respect to gravity deformations for any spherically symmetric spacetime, hence explaining the insensitivity of $\vartheta$ with respect to gravity deformations. 

The remaining uncertainty in $\vartheta$ derives from the functional form of the invariant intensity as a function of impact parameter normalized to the critical one.  This functional form depends more strongly on the emission model than the spacetime metric. In fact, we find that the main effect of deforming gravity is to move the location of its bright peak; the overall shape of the emission profile is nearly unchanged by gravity deformations.  It can be significantly changed by altering the emission profile. 

We conclude that the characteristic radius of emission of a non-spinning black hole is changed significantly by altering the astrophysical model, and is relatively insensitive to the GR deformation in both the sGB theory and in the RZ metric, as suggested by \cite{gralla}. The ratio $\vartheta$ does depend on the spacetime metric, albeit very weakly, so if the astrophysical emission model is sufficiently well-understood, controlled tests of GR with EHT observations are possible.

This paper is structured as follows. In \textsection~\ref{sec:overview} we outline the framework of our calculation; in \textsection~\ref{sub:grav} we discuss the spacetimes used and in \textsection~\ref{sub:astro} we outline the astrophysical components of our model, namely, the accretion models in \textsection~\ref{subsub:acc} and the emissivity model in \textsection~\ref{subsub:emis}. In \textsection~\ref{sec:radin} we show how $\vartheta$ changes as we modify the astrophysical model and gravity theory in the radial freefall (cold) accretion scenario. In \textsection~\ref{sec:bondi}, we do the same as in \textsection~\ref{sec:radin}, but in the Bondi accretion model. We also provide a derivation of the Bondi flow in an arbitrary spherically symmetric spacetime in the Schwarzschild coordinate chart. In \textsection~\ref{sec:phenom}, we compute a semi-analytic model for the intensity profile $I(b)$ in the cold  accretion model. We further use this result to put our numerical findings on firm theoretical grounds. We conclude with a discussion in \textsection~\ref{sec:discuss}. Throughout we set $G = c = 1$. Greek indices represent spacetime indices ranging from 0 to 3, whereas Latin indices represent purely spatial indices ranging from 1 to 3. The spacetime signature throughout is $(-,+,+,+)$. 

\section{Theoretical framework} \label{sec:overview}

EHT observations detect photons emitted by plasma close to the event horizon. It is natural to decompose our model into a gravitational component and an astrophysical component. The gravitational component is simply a model for the black hole spacetime (the weak-equivalence principle is assumed to hold). The astrophysical model includes both a flow model (cold [free fall] model or warm [Bondi] model) and an emissivity model (here a power-law in radius). These are described in \textsection~\ref{sub:grav} and \textsection~\ref{sub:astro} respectively. 

Along the way we will need to compute intensity $I$ (the specific intensity, usually denoted $I_\nu$) as a function of impact parameter $b$ in the frame of our telescope. To compute $I(b)$, we use a numerical ray tracing code where null rays are cast backwards in time from the telescope. Integration is halted when a null ray either reaches the event horizon or has gone past the black hole to a radius that is sufficiently large that the emissivity is negligible ($r > 30 \ M$ in this paper). The radiative transfer equation is then integrated forward along the ray to the telescope. 

The invariant, unpolarized radiative transfer equation is 
\begin{equation}
    \frac{d \iI}{d\lambda} = \frac{j_\nu}{\nu^2} - (\nu\alpha_\nu) \iI.
\end{equation}
where $\lambda$ is the affine parameter along a null geodesic.  Here ``invariant'' means frame-independent.  $\iI = I/\nu^3$ is the invariant intensity, proportional to the photon phase space density, $\nu$ is the frequency in the plasma frame, $j_\nu$ is the emissivity that we will express in terms of the invariant emissivity $J := j_\nu / \nu^{2}$, and $\alpha_\nu$ is the absorptivity.  We will assume zero optical depth and set $\alpha_\nu = 0$.

\par 

With these approximations the transfer equation is
\begin{equation}
    \frac{d \iI}{d\lambda} = J. \label{eq:radtrans}
\end{equation}
The null ray is a solution to the geodesic equation, and $k^\mu = dx^\mu/d\lambda$ is the photon wavevector; $k^\mu k_\mu = 0$.  The frequency is evaluated using $\nu = -k^\mu u_\mu$ where $u^\mu$ is the plasma four-velocity.  The intensity on the sky at the observer is $I = \nu^3 \iI$.  Evidently if $j_\nu$ is specified then $J = j_\nu/\nu^2$ depends on the plasma four-velocity. 

\subsection{Spherically symmetric spacetimes} \label{sub:grav}

We consider spherical accretion on to a spherically symmetric astrophysical black hole of mass \textit{M}. We write the generic spacetime metric of such an object as
\begin{equation} 
ds^{2} = g_{tt} dt^{2} + g_{rr} dr^{2} + r^{2} \left( d\theta^{2} + \sin^{2}\theta d\phi^{2}\right), \label{eq:genmetric}
\end{equation}
where $\{ t, r, \theta, \phi \}$ are the canonical Schwarzschild coordinates. We use $r_{g} := MG/c^{2}$ as our measure of distance, and set $M = 1$ for simplicity.

Two features of the spacetime are of particular importance: the radius of the photon orbit, $r_{ph}$, and $b_c$, the critical impact parameter for rays with minimum radius $r_{ph}$. Computing these quantities in a spherically symmetric spacetime is straightforward.  Consider null rays orbiting a black hole, without loss of generality in the equatorial plane, at fixed radius. Then along the ray
\begin{equation}
    g_{tt} dt^{2} + g_{\phi\phi} d\phi^{2} = 0, \label{eq:lineel}
\end{equation}
where $ds^{2} = 0$ since the rays are null. Then
\begin{equation}
   \dfrac{d\phi}{dt} = \sqrt{-\dfrac{g_{tt}}{g_{\phi\phi}}}. \label{eq:dphidtone}
\end{equation}
We can find this quantity directly using the geodesic equation; evaluating the geodesic equation for the radial coordinate,  
\begin{equation}
    \dfrac{d\phi}{dt} = - \dfrac{\Gamma^{r}_{tt}}{\Gamma^{r}_{\phi\phi}}, \label{eq:christoffs}
\end{equation}
where $\Gamma^{\mu}_{\nu\lambda}$ are the Christoffel symbols. Equating \eqref{eq:dphidtone} and \eqref{eq:christoffs} gives $r_{ph}$. Noting that rays on the photon sphere have $d\phi/dt = b_{c}^{-2}$, evaluating \eqref{eq:dphidtone} at $r = r_{ph}$ gives $b_{c}$. 

In GR, the metric functions $g_{tt}$ and $g_{rr}$ for a stationary, spherically symmetric black hole are
\begin{equation}
- g_{tt}^{GR} = \dfrac{1}{g_{rr}^{GR}} =  1 - \dfrac{2}{r} .
\end{equation} 
Applying the prescription given above for computing $r_{ph}$ and $b_{c}$, we recover the well-known results~\citep{bardeen}
\begin{equation}
    r_{ph}^{GR} = 3, \qquad b_{c}^{GR} = \sqrt{27}, \label{eq:bcritGR}
\end{equation}
as expected. We discuss the metric functions used for scalar Gauss -- Bonnet gravity and the Rezzolla and Zhidenko spacetime in \textsection~\ref{subsub:scalar} and \textsection~\ref{subsub:rezz}, respectively, as well as $r_{ph}$ and $b_c$ in both theories.

\begin{figure*}
\centering
\includegraphics[width=\textwidth]{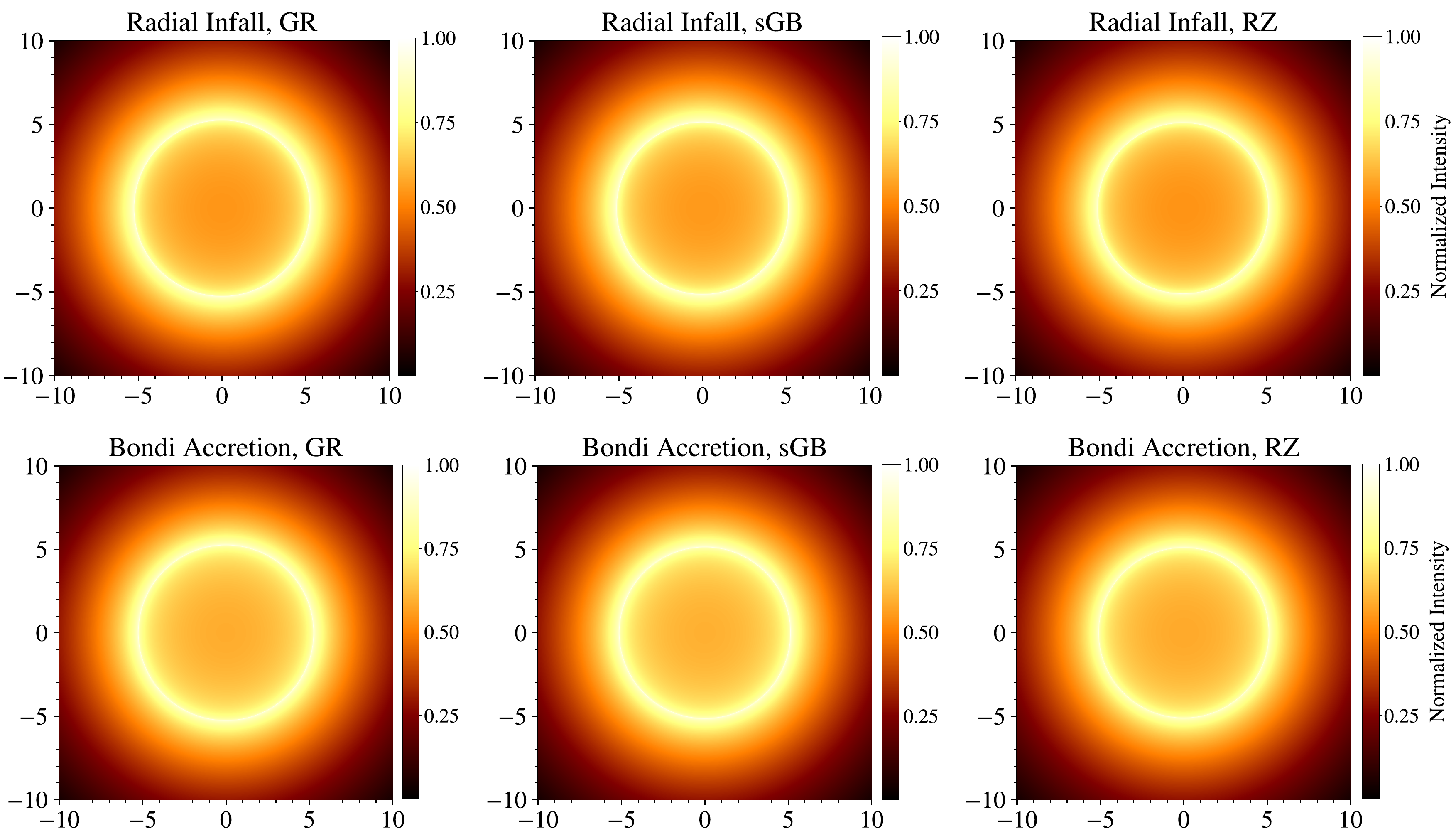}
\caption{\textit{(Left panels)}. GR black hole shadows in the radial infall (top) and Bondi accretion (bottom) cases. \textit{(Middle panels)}. sGB black hole shadows in the radial infall (top) and Bondi accretion (bottom) cases with the coupling constant $\zeta$ set to its max value in our simulations, $\zeta = 0.175$. \textit{(Right Panels).} RZ black hole shadows for the radial infall (top) and Bondi accretion (bottom) scenarios with the deformation parameter $a_{1} = 0.2$. For the radial infall images, $\alpha = 6$. For the Bondi images, $\alpha = 6$, $\tilde{R} = 10$, and $\gamma = 5/3$. The logarithmic intensity in all images is normalized. Notice that the shape and overall brightness profile of the black hole shadow is not noticeably changed between theories.}
\label{fig:images}
\end{figure*}

\subsubsection{Scalar Gauss -- Bonnet gravity} \label{subsub:scalar}

The first non-Schwarzschild spacetime we consider is a solution to a particular modified theory of gravity, scalar Gauss-Bonnet gravity, which is a well-motivated member of a class of modified theories with actions that include all possible algebraic, quadratic curvature scalars. As shown in \cite{yunesstein}, spherically symmetric and static solutions can be found analytically and in closed-form within the {\emph{small-coupling approximation}}, where one expands the modified field equations in a dimensionless form of the theory's coupling parameter $\zeta$; the resulting modified spacetime reduces to the Schwarzschild metric smoothly as $\zeta$ vanishes. Therefore, these solutions describe continuous deformations of the Schwarzschild metric, whose strength is mediated by a dimensionless coupling parameter \citep{campbell,kanti,yunesstein}.

We appropriate the solutions to the modified field equations in sGB using the notation of \cite{yunesstein}, which are valid to linear order in $\zeta$, as 
\begin{align}
g_{tt}^{sGB}  & = - f(r) \left(1 + \dfrac{\zeta}{3 r^{3} f(r)} h(r)\right), \label{eq:gttnoresum} \\
g_{rr}^{sGB} & = \dfrac{1}{f(r)} \left( 1 - \dfrac{\zeta}{r^{2}f(r)} k(r) \right), \label{eq:grrnoresum}
\end{align}
where
\begin{align}
h(r) & := 1 + \dfrac{26}{r} + \dfrac{66}{5r^{2}} + \dfrac{96}{5r^{3}} - \dfrac{80}{r^{4}}, \\
k(r) & := 1 + \dfrac{1}{r} + \dfrac{52}{3r^{2}} + \dfrac{2}{r^{3}} + \dfrac{16}{5r^{4}} - \dfrac{368}{3r^{5}}, \\
f(r) & := 1 - \dfrac{2}{r},
\end{align}
with the following caveat. Our intention is to numerically integrate the radiative transfer equation in sGB gravity from the black hole event horizon to the location of our telescope, effectively infinitely far away. However, \eqref{eq:gttnoresum} and \eqref{eq:grrnoresum} are only valid over the GR Schwarzschild coordinate ranges, i.e., $r > 2$ \citep{yunesstein}. This poses a complication for numerical integration, as it was shown in \cite{yunesstein} that the horizon location in sGB gravity, $r_{H} :=  2 - (49/40)\zeta$, is changed such that $r_{H} < 2$ for all $\zeta$. This means that during a numerical integration, starting at $r = r_{H}$, one would encounter a coordinate singularity at $r = 2$ if \eqref{eq:gttnoresum} and \eqref{eq:grrnoresum} are taken to be the metric components verbatim.
\par 
We overcome this complication by performing a resummation to extend the coordinate range of the radial coordinate to $r > r_{H}$. To do this, we define a new function $\bar{f}(r)$ as 
\begin{equation}
\bar{f}(r) := 1 - \dfrac{r_{H}}{r}, \label{eq:fbar}
\end{equation}
and assume that the resummed metric components, $\bar{g}_{tt}$ and $\bar{g}_{rr}$, have the form of 
\begin{align}
\bar{g}^{sGB}_{tt} & = - \bar{f}(r) \left( 1 + \dfrac{\zeta}{3 r^{3} \bar{f}(r)} h(r) + \mathcal{H}(r) \right), \label{eq:gttresum} \\
\bar{g}^{sGB}_{rr} & = \dfrac{1}{\bar{f}(r)} \left( 1 - \dfrac{\zeta}{r^{2}\bar{f}(r)} k(r) + \mathcal{K}(r) \right), \label{eq:grrresum} 
\end{align}
where ${\cal{H}}$ and ${\cal{K}}$ are undetermined functions of radius only. By demanding that \eqref{eq:gttnoresum} (\eqref{eq:grrnoresum}, resp.) is exactly equal to \eqref{eq:gttresum} (\eqref{eq:grrresum}, resp.) to first order in $\zeta$, we find that,
\begin{equation}
\mathcal{H}(r) = -\mathcal{K}(r) = - \dfrac{49 \zeta}{40 r \bar{f}(r)}, \label{eq:sksub}
\end{equation}
to first order in $\zeta$. Using \eqref{eq:sksub} in \eqref{eq:gttresum} and \eqref{eq:grrresum} completes the resummation, hence removing all coordinate pathologies at $r = 2$, as desired. 

An added benefit of using the resummed metric is that we may now treat this metric as exact for the remainder of our work, making it particularly robust for numeric and analytic calculations. Note that, however, \eqref{eq:gttresum} and \eqref{eq:grrresum} are technically only valid for sufficiently small values of the coupling parameter $\zeta$; more specifically, one can show that $0 < \zeta < 24/131$ by demanding the metric be Lorentzian, i.e., demanding det$(g) < 0$, for all $r > r_{H}$. If one wished to use the sGB metric with higher values of the coupling parameter, one would then need to resum the higher-order-in-$\zeta$ solution to sGB gravity~\citep{maselli}. Of course, the results of doing so will be very similar to what we find here for sufficiently small $\zeta$.

\par 

Immediately one may compute the correction to the location of the photon orbit and critical impact parameter for null rays in sGB gravity via the prescription given in \textsection~\ref{sub:grav}. Computing these quantities to first order in the coupling constant, we find that the photon orbit is altered such that 
\begin{equation}
r_{ph}^{sGB} = 3 \left( 1 - \dfrac{961}{2430} \zeta \right), \label{eq:rphot}
\end{equation}
and the critical impact parameter is shifted such that
\begin{equation}
b_{c}^{sGB} = \sqrt{27}\left( 1 - \dfrac{4397}{21870} \zeta \right). \label{eq:bcritsGB}
\end{equation}
Note that both of these results are linear corrections to the well known results in GR, as expected from effective field theory. Moreover, the numerical coefficients of the sGB correction (the numbers multiplying $\zeta$) are $\sim 0.4$ and $\sim 0.2$, and thus of ${\cal{O}}(1)$ and not unnaturally small.

\subsubsection{Rezzolla and Zhidenko spacetime} \label{subsub:rezz}

The second modified spacetime we consider will be described by the Rezzolla and Zhidenko (RZ) metric. The framework that leads to the RZ metric is significantly different from how one obtains the sGB-modified metric of the previous section. In sGB gravity, there is an explicit action, and therefore, one can find explicit solutions to the modified field equations, for example in the small coupling limit. The RZ metric, on the other hand, is \textit{not} formulated from a given action, but rather it is constructed by considering general deformations to well-known GR solutions (in the case of spherical symmetry, deformations of the Schwarzschild spacetime), which are parameterized by a family of undetermined parameters~\citep{Rezzolla:2014mua}; metrics that are constructed in this way are often referred to as ``bumpy" black hole metrics~\citep{bumpy}. Using the RZ metric therefore allows us to investigate the impact deformations of GR would have on a given calculation without specifying an exact theory of interest, whose spherically-symmetric solution is the RZ metric.
\par 
However, there are certain limitations that come with using a ``bumpy" metric. Indeed, mapping the RZ metric to solutions in known modified gravity theories (such as sGB gravity) typically requires many bumpy parameters in the bumpy metric, which render analytical calculations intractable in full generality. Therefore, this approach has been followed using only the leading-order terms of the parametrization, and any results derived from it must therefore be taken with great care~\citep{sebas}. Here we use the RZ metric with the lowest unconstrained deformation parameter that enters into our calculation, $a_{1}$, and neglect all others. This metric therefore does not map to solutions of any known modified gravity theory. 

\par 

The RZ line element takes a simple functional form when only one bumpy parameter is considered. The parametrization assumes the metric functions $-g^{RZ}_{tt}=1/g^{RZ}_{rr}=N^2(x)$, where $N(x)$ is written in terms of a compactified radial coordinate 

\begin{equation}
x := 1 - \frac{r_{H}}{r},
\end{equation}

so that $x=0$ corresponds to the location of the event horizon ($r_H=2$), while $x=1$ corresponds to spatial infinity. In terms of the compactified radial coordinate, this function is written as~\citep{Rezzolla:2014mua}

\begin{equation}
N^{2}(x):=x A(x),
\end{equation}
where the function $A(x)$ is a Taylor-expansion about $x=1$ and we here consider it up to $\mathcal{O}[\left( 1 - x\right)^{4}]$ with only one bumpy parameter. Explicitly, this function takes the following simple form 

\begin{equation}\label{metricthermal}
A(x):= 1+a_1(1-x)^3.
\end{equation}

The bumpy parameter $a_{1}$ characterizes the magnitude of the non-Schwarzschild deformation; when $a_{1}=0$ the metric reduces exactly to Schwarzschild. The RZ metric is reasonable only for small deformations away from Schwarzschild, i.e.~$|a_{1}| \ll 1$, because otherwise, certain pathologies may arise in the spacetime, such as the loss of Lorentzian signature. We here take $|a_{1}| < 0.2$ in all of our calculations. 

With this metric, one can now easily calculate the properties of null geodesics. More specifically, the location of the photon sphere is shifted such that
\begin{equation}
r_{ph}^{RZ} = 3 \left( 1  - \dfrac{12}{81} a_{1} \right),
\end{equation}
which moves the critical impact parameter to 
\begin{equation}
b_{crit}^{RZ} = \sqrt{27} \left( 1 - \dfrac{4}{27}a_{1} \right). \label{eq:bcritRZ}
\end{equation}
In the next subsection, we introduce our astrophysical model that complements our gravity theories when computing $\vartheta$.

\subsection{Astrophysical model} \label{sub:astro}

Finding $\vartheta$ requires the specification of an accretion flow model and an emissivity model. We allow for free parameters in both the accretion and emissivity models so that we can assess the sensitivity of $\vartheta$ to changes in the astrophysical model. This, in turn, will allow us to compare this astrophysical sensitivity to the sensitivity of $\vartheta$ with respect to variations in the gravity theory, as represented through the variation of the magnitude of the deformation parameter $\zeta$ in sGB and $a_{1}$ in RZ. 

\subsubsection{Accretion models} \label{subsub:acc}

We choose to calculate $\vartheta$ using the Bondi spherical accretion model in both the zero pressure limit (i.e., radial free fall) and in the nonzero pressure case. Given that the flow of radially infalling plasma only depends on the spacetime metric, this accretion model serves as a toy model to gain intuition for the behavior of $\vartheta$ in response to changing our gravity theory \citep{narayan2019}. Additionally, it is simple enough to be analytically tractable, allowing us to compute the intensity profile semi-analytically; see \textsection~\ref{sec:phenom} for the full derivation. 

\par 

The cold accretion model does not depend on any accretion parameters. We also study a warm accretion or Bondi flow model. This model has four accretion parameters: the black hole mass, \textit{M}, the accretion rate, $\dot{M}$, the radius at which the flow passes through the critical point, $\tilde{R}$, and the adiabatic index, $\gamma$ \citep{shapiro}.  The warm accretion model depends on these parameter {\em and} the spacetime metric. This model allows us to test the sensitivity of $\vartheta$ to both changing the gravity theory and the plasma flow profile.
 
\subsubsection{Emissivity} \label{subsub:emis}

We specify a model for the emissivity $j_\nu$ by considering the following. We desire a model for the emissivity that is smooth for the entire coordinate domain, is spherically symmetric, is largest near the black hole, and is simple enough to be analytically tractable. Hence, we invoke a power law spatial dependence of $j_\nu$ on the radial coordinate, as was done in \cite{narayan2019}, such that
\begin{equation}
j_{\nu} = j_{0} \left(\dfrac{r_{0}}{r}\right)^{\alpha}, \label{eq:jnu}
\end{equation}
where $r_{0}$ and $j_{0}$ are positive constants. In all numerical work we set $r_{0} = j_{0} = 1$. Notice that $j_\nu$ is independent of frequency; in EHT models this would correspond to emission close to the peak of the synchrotron emissivity.  We allow $\alpha$ to vary over $5 < \alpha \leq 8$; $\alpha > 5$ is required for the convergence of \eqref{eq:rchardef}.

\section{Toy model: cold accretion} \label{sec:radin}

\begin{figure}
\includegraphics[width=\columnwidth]{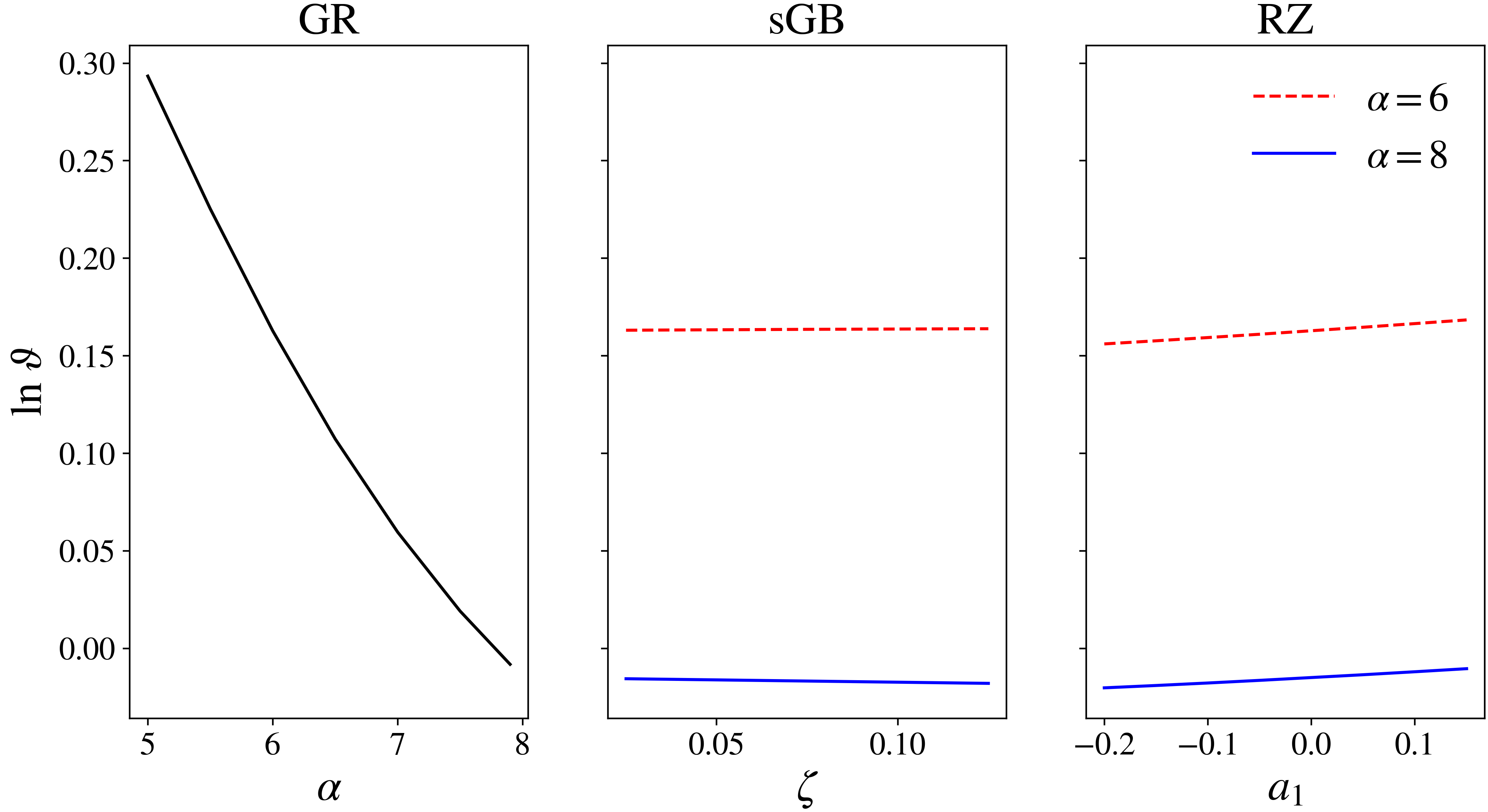}
\caption{\textit{(Left).} Shown is a plot of $\ln\vartheta$ against the power law scaling of the frame dependent emissivity, $\alpha$, calculated using the radial infall model in GR. In the remaining two panels we plot $\ln\vartheta$ against the sGB coupling parameter $\zeta$ and RZ deformation parameter $a_{1}$, for fixed powers of the emissivity as $\alpha=6$ and $\alpha = 8$, respectively, in the same accretion model. There is significant change in the value of $\ln\vartheta$ when we vary $\alpha$, but minimal variation when $\zeta$ or $a_{1}$ are changed.}
\label{fig:radinmeanrad}
\end{figure}

We first consider the simple model of cold, optically thin infall onto a black hole. The gas is assumed to be at rest at large $r$; together with spherical symmetry this implies that $u_\mu = (-1, u_r, 0, 0)$.  Normalization of four-velocity,  $u_{\mu}u^{\mu} = -1$, implies $u_{r} = \sqrt{- g_{rr}(1 + g^{tt})}$. The flow of the gas depends on the metric, and thus the gravity theory. 

Using this model for $u_\mu$, we compute $I(b)$ using a ray-tracing code for $5 < \alpha \leq 8$, $0 \leq \zeta < 24/131$, and $|a_{1}| < 0.2$.  Then $I(b)$ is used in \eqref{eq:rchardef} to compute the characteristic radius of emission, which in tandem with \eqref{eq:bcritGR}, \eqref{eq:bcritsGB} and \eqref{eq:bcritRZ} allows us to find $\vartheta$ in each metric considered.

The top panels in Fig.~\ref{fig:images} show three example synthetic EHT images, one in each spacetime, using the cold accretion model. Evidently the size of the black hole shadow is not noticeably changed by the spacetime metric, and the overall shape of the shadow is consistent across all three examples.

\par 

Our findings from the procedure described above are summarized in Figure~\ref{fig:radinmeanrad}, where we plot $\ln(\vartheta$) against $5 < \alpha \leq 8$, $0 \leq \zeta < 24/131$, and $-0.2 < a_{1} < 0.2$. It is clear from this figure that, in this toy model, $\vartheta$ depends strongly on $\alpha$ and hardly at all on $\zeta$ or $a_{1}$. Indeed, computing a linear regression of the above data, reveals that $\vartheta$ scales with $\alpha$ as $\vartheta_{GR} \sim \mathrm{exp}\left( -0.1 \alpha\right) \approx 1 - 0.1 \alpha + {\cal{O}}(\alpha^2)$ in GR, whereas in a non-Schwarzschild spacetime  $\vartheta_{sGB}/\vartheta_{GR} \sim \mathrm{exp}(0.008 \zeta) \approx 1 + 0.008 \zeta + {\cal{O}}(\zeta^2)$ and $\vartheta_{RZ}/\vartheta_{GR} \sim \mathrm{exp}(0.04 a_{1}) \approx  1 + 0.04 a_1 + {\cal{O}}(a_1^2)$ for $\alpha = 6$. Notice that as the deformation parameters $\zeta, a_{1} \to 0$ we recover the GR value for $\vartheta$, as expected.

\par 

Since $\zeta, a_{1} \sim {\cal{O}}(10^{-1})$ and $\alpha \sim \mathcal{O}(1)$, the impact of  non-Schwarzschild deformations on $\vartheta$ are, at best, two orders of magnitude smaller than the impact of changing $\alpha$. This is consistent with~\cite{gralla}; the ratio of the characteristic radius of emission to the critical impact parameter is far more sensitive to the astrophysical emission model than the spacetime metric, and thus, non-GR deviations. 

Our results also suggest, however, that the calibration of EHT images using GR based simulations will not be affected by non-GR metric deformations. This, in turn, implies that, provided one has a detailed understanding of the accretion model, one could, in principle, test GR using EHT data. This is because the size of the observed shadow depends on the characteristic radius, $r_{char}$, and non-Schwarzschild corrections in this quantity are much larger than those in the ratio $\vartheta$. Given that this was only a toy model, it would be beneficial to see if this trend remains the same in more sophisticated accretion models. 

\section{Warm accretion model} \label{sec:bondi}

In this section, we upgrade our previous model by considering Bondi or warm accretion, where radial pressure gradients in the inflow can affect the radial velocity profile.  The Bondi model, in addition to being more realistic, offers additional astrophysical parameters that we can use to test the sensitivity of $\vartheta$ to the astrophysical model.
\par 
Before moving forward we need to find warm accretion plasma velocity in a general spherically symmetric spacetime.  Our derivation follows \cite{shapiro}, who used free metric functions in GR in their derivation. Notice that a similar derivation with a different metric parameterization appears in \cite{chaverrasarbach}, and the generalization of Bernoulli's theorem can be found in \cite{genradial}. We provide the full derivation here for clarity.  Notice that we use the {\em test fluid} approximation and neglect the self-gravity of the accretion flow. This is an excellent approximation, especially for the low accretion rates typical of EHT sources.

\subsection{Warm accretion velocity profile in general spherically symmetric spacetimes}

Consider spherical Bondi accretion in a general spherically symmetric background spacetime in the Schwarzschild coordinate chart. We assume the spacetime metric in \eqref{eq:genmetric} where $(g_{tt}, g_{rr}) \in \mathcal{C}^{1}(\mathbb{R})$ are first differentiable functions of radius only, and are continuous over the entire radial coordinate domain. Further, we assume that $g^{rr} = 1/g_{rr} = 0$ defines an event horizon, with the greatest positive solution at $ r = r_{H}$. Lastly, we assume the spacetime is asymptotically flat, i.e., $\lim_{r\to \infty} g_{\alpha\beta} = \eta_{\alpha\beta}$, where $\eta_{\alpha\beta}$ is the Minkowski metric.

Consider a fluid with total density $\rho$, proper rest-mass density $\rho_{0}$ and internal energy density $\varepsilon$, such that $\rho = \rho_{0} + \varepsilon$. We assume a Gamma law equation of state, with pressure $P  =(\gamma - 1)\varepsilon$. If the flow is isentropic then this implies $P = \kappa \rho_{0}^{\gamma}$, where $\kappa$ is a constant \citep{shapiro}.  The stress energy tensor is
\begin{equation}
T^{\alpha\beta} = \rho u^{\alpha} u^{\beta} + P \Delta^{\alpha\beta}, \label{eq:SEtensor}
\end{equation}
where $u^{\mu} = \mathrm{col}(u^{t}, u^{r}, 0,0)$ is the four velocity of the gas, $\Delta^{\alpha\beta} := g^{\alpha\beta} + u^{\alpha}u^{\beta}$ is the projector orthogonal to the fluid flow $u^{\alpha}$ and \textit{P} $= P(r)$ is the pressure, assumed to be a function of radius only. 

The law of baryon conservation and conservation of energy-momentum are
\begin{align}
\nabla_{\alpha}\left( \rho_{0} u^{\alpha} \right) & = 0, \label{eq:baryoncons} \\
\nabla_{\alpha} T^{\beta\alpha} & = 0, \label{eq:SEcons} 
\end{align}
respectively. From \eqref{eq:baryoncons},
\begin{equation}
\dfrac{\rho_{0}'}{\rho_{0}} + \dfrac{u'}{u} + \Sigma = 0, \label{eq:baryF}
\end{equation}
with $u := u^{r}$, $\Sigma := \left( \sqrt{-g} \right)' / \sqrt{-g}$.  Here $g := \mathrm{det} (g_{\mu\nu})$ is the determinant of the metric, and $' := d/dr$. In deriving \eqref{eq:baryF} we assumed the metric is time independent.
\par 
Expanding \eqref{eq:SEcons} yields
\begin{equation}
(\rho + P) u^{\beta} \nabla_{\beta} u^{\alpha} + \Delta^{\alpha\beta} \partial_{\beta} P = 0, \label{eq:secons1}
\end{equation}
and evaluating the radial component of \eqref{eq:secons1} results in 
\begin{equation}
u u' - \sigma (1 + g_{rr} u^{2} ) + \Gamma^{r}_{rr} u^{2} = - c_{s}^{2} (g^{rr}+ u^{2} ) \dfrac{\rho_{0}'}{\rho_{0}}, \label{eq:seconsF}
\end{equation}
where $\sigma := \Gamma^{r}_{tt} / g_{tt}$ and $c_{s}^{2} := dP/d\rho$ is the sound speed. In deriving \eqref{eq:seconsF} we used the definition of $c_{s}^{2}$ and the first law of thermodynamics in the form
\begin{equation}
P' = (\rho + P) c_{s}^{2} \dfrac{\rho_{0}'}{\rho_{0}}. \nonumber
\end{equation}

\subsection{Solution at the critical radius}

We can simultaneously solve \eqref{eq:baryF} and \eqref{eq:seconsF} to eliminate $u'$ and $\rho_{0}'$, giving
\begin{equation}
\dfrac{u'}{u} = \dfrac{\mathcal{D}_{1}}{\mathcal{D}}, \qquad \dfrac{\rho_{0}'}{\rho_{0}} = - \dfrac{\mathcal{D}_{2}}{\mathcal{D}}, \nonumber
\end{equation}
where
\begin{align}
\mathcal{D}_{1} & := c_{s}^{2} \Sigma (u^{2} + g^{rr}) + \sigma (1 + g_{rr}u^{2}) - \Gamma^{r}_{rr} u^{2}, \label{eq:d1} \\
\mathcal{D}_{2} & := \sigma (1 + g_{rr}u^{2}) + u^{2} (\Sigma - \Gamma^{r}_{rr} ), \label{eq:d2} \\
\mathcal{D} & := u^{2} - c_{s}^{2} (g^{rr} + u^{2}). \label{eq:d}
\end{align}

We will now demonstrate that for an equation of state obeying the causality condition $c_{s}^{2} < 1$, the flow must pass through a critical point outside the horizon radius $ r = r_{H}$. First consider \eqref{eq:d} in the limit where $r \to \infty$. Then, since the spacetimes considered here are all asymptotically flat, we have that 
\begin{equation}
\lim_{r\to\infty} \mathcal{D} \sim  - c_{s}^{2} < 0, \nonumber
\end{equation}
since $u^{2} \to 0$ and $g^{rr} \to 1$ when $r \to \infty$. Now considering the limit where $r \to r_{H}$, we find that 
\begin{equation}
\lim_{r \to r_{H}} \mathcal{D} = u^{2} (1 - c_{s}^{2} ) > 0, \nonumber
\end{equation}
where the inequality follows from the causality condition. Therefore, by continuity, we must pass through a point $r = \tilde{R}$ where $\mathcal{D} = 0$. 

For the flow to be regular at this critical point, we must have $\mathcal{D}_{1} = \mathcal{D}_{2} = 0$ at $r = \tilde{R}$. In what follows, quantities with tildes indicate evaluation at the critical radius $r = \tilde{R}$. Simultaneously solving \eqref{eq:d1} and \eqref{eq:d2}, the speed of the gas and the sound speed at the critical point are given by
\begin{align}
\tilde{u}^{2} & = \dfrac{\sigma}{\Gamma^{r}_{rr} - \Sigma - \sigma g_{rr}} \Bigg\vert_{r = \tilde{R}}, \label{eq:utilde} \\
\tilde{c}_{s}^{2} & = \dfrac{\sigma}{\sigma + g^{rr} \left( \Gamma^{r}_{rr} - \Sigma - \sigma g_{rr}\right)}\Bigg\vert_{r = \tilde{R}}. \label{eq:atilde} 
\end{align}
This concludes the analysis at the critical radius. 

\subsubsection{Conservation equations}

Using the definition of $\Sigma$, \eqref{eq:baryF} can be directly integrated as  
\begin{equation}
4 \pi u \rho_{0} \sqrt{-g} = \dot{M}, \label{eq:masscons}
\end{equation}
where $\dot{M}$ is the accretion rate. 

Then \eqref{eq:seconsF} implies
\begin{equation}
c_{s}^{2} \dfrac{\rho_{0}'}{\rho_{0}} = \dfrac{\rho_{0}}{\rho + P} \left( \dfrac{\rho + P}{\rho_{0}} \right)', \label{eq:asquaredsub}
\end{equation}
using the first law of thermodynamics and the equation of state. Starting with  \eqref{eq:asquaredsub} and expanding the Christoffel symbols, after some algebra \eqref{eq:seconsF} can be rewritten as 
\begin{equation}
\dfrac{1}{2} \dfrac{\left( 1 + u^{2}/g^{rr} \right)'}{1 + u^{2}/g^{rr}} + \dfrac{1}{2} \dfrac{g_{tt}'}{g_{tt}} +  \dfrac{\rho_{0}}{\rho + P} \left( \dfrac{\rho + P}{\rho_{0}} \right)' = 0, \nonumber
\end{equation}
which can be integrated to yield the generalized Bernoulli equation
\begin{equation}
| g_{tt} | \left( 1 + \dfrac{u^{2}}{g^{rr}}\right)  \left( \dfrac{\rho + P}{\rho_{0}} \right)^{2} = C, \label{eq:relbernulgen}
\end{equation}
where $C$ is a constant.

\subsubsection{Solving for velocity and density profiles}

Let us now cast the expressions above into a form suitable for finding $u(r)$ for some choice of parameters $(M, \dot{M}, \tilde{R}, \gamma)$. We can first use the definition of $c_{s}^{2}$, the equation of state, and the first law of thermodynamics to show (see \cite{shapiro} for a full derivation) that
\begin{equation}
\dfrac{\rho + P}{\rho_{0}} = 1 + \dfrac{c_{s}^{2}}{\gamma - 1 - c_{s}^{2}}. \label{eq:rhoPrho0sub}
\end{equation} 
Using \eqref{eq:rhoPrho0sub} in \eqref{eq:relbernulgen} gives
\begin{equation}
| g_{tt} | \left( 1 + \dfrac{u^{2}}{g^{rr}}\right) \left( 1 + \dfrac{c_{s}^{2}}{\gamma - 1 - c_{s}^{2}} \right)^{2} = C. \label{eq:subbbbed}
\end{equation}
We now note that the left-and side of \eqref{eq:subbbbed} is in terms of quantities that are known at the critical point: \textit{u}, $c_{s}$, and \textit{r}. Therefore, we can evaluate \eqref{eq:subbbbed} at $r = \tilde{R}$, which determines the Bernoulli constant as 
\begin{equation}
C \equiv \left( \dfrac{\gamma - 1}{\gamma - 1 - \tilde{c}_{s}^{2}} \right)^{2} \left( 1 + \dfrac{\tilde{u}^{2}}{\tilde{g}^{rr}} \right) | \tilde{g}_{tt}|.
\end{equation}
Using a procedure similar to that used to obtain \eqref{eq:rhoPrho0sub},
\begin{equation}
\dfrac{\rho + P}{\rho_{0}} = 1 + \dfrac{\gamma}{\gamma - 1} \kappa \rho_{0}^{\gamma - 1}. \label{eq:almostthere}
\end{equation}
Setting \eqref{eq:almostthere} equal to \eqref{eq:rhoPrho0sub},
\begin{equation}
\kappa = \dfrac{(\gamma  - 1)\tilde{c}_{s}^{2}}{\gamma \tilde{\rho_{0}}^{\gamma - 1} (\gamma  - 1 - \tilde{c}_{s}^{2})}, 
\end{equation}
where we have used
\begin{equation}
\tilde{\rho_{0}} = \dfrac{\dot{M}}{4 \pi \tilde{u} \sqrt{-\tilde{g}}}
\end{equation}
by evaluating \eqref{eq:masscons} at the critical point. Finally, using \eqref{eq:masscons} and \eqref{eq:almostthere}, we can write \eqref{eq:subbbbed} in terms of \textit{u} only as
\begin{equation}
|g_{tt}| \left( 1 + \dfrac{\gamma\kappa}{\gamma - 1} \left( \dfrac{\dot{M}}{4\pi u \sqrt{-g}} \right)^{\gamma - 1} \right)^{2} \left( 1 + \dfrac{u^{2}}{g^{rr}}\right) = C, \label{eq:ufinal}
\end{equation}
as originally desired.  This is a nonlinear equation for $u(r)$ that can be solved using a numerical root-finding scheme.  

\begin{figure}
\includegraphics[width=\columnwidth]{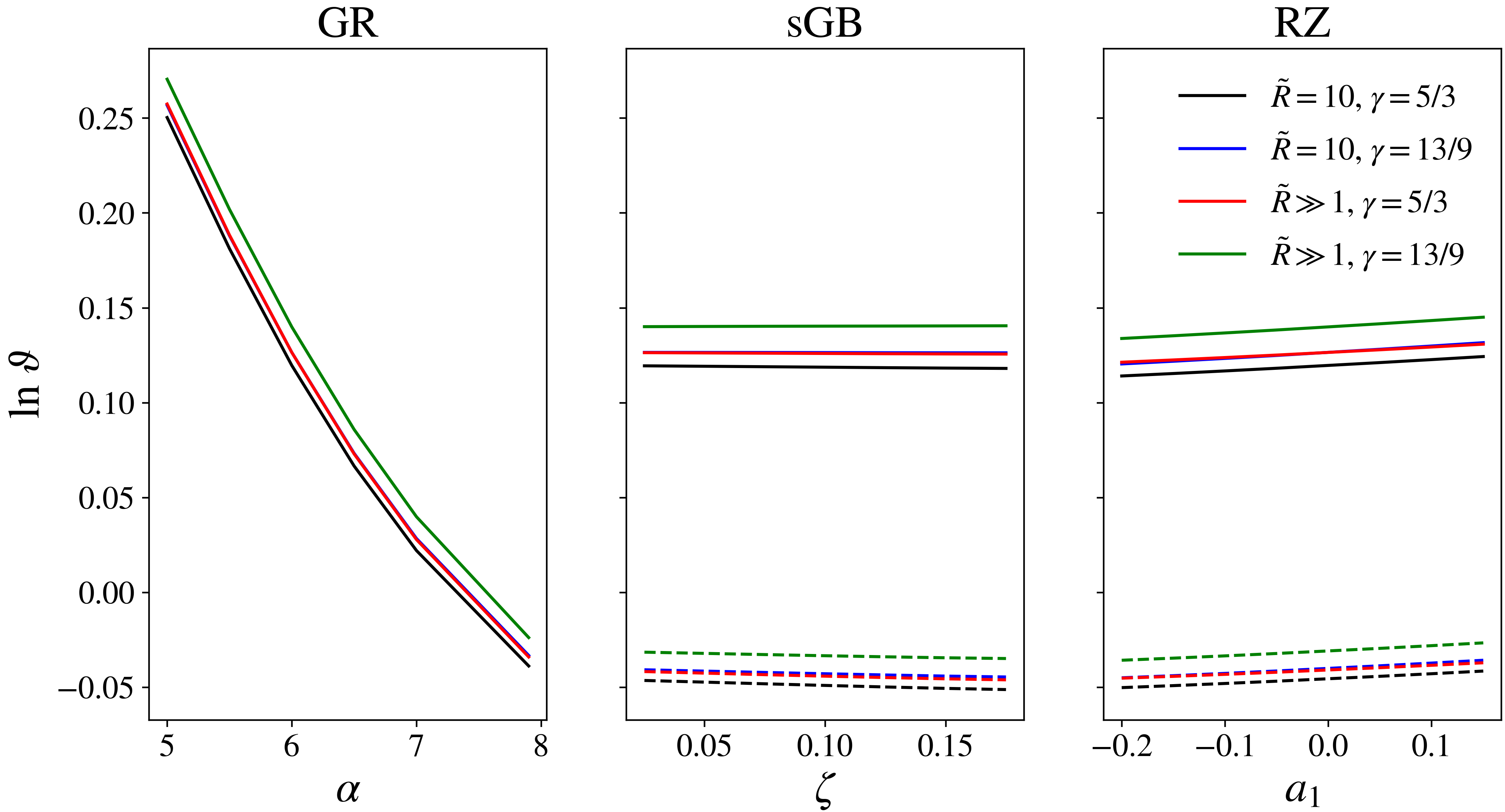}
\caption{\textit{(Left).} Shown is the plot of $\ln\vartheta$ against the power law scaling of the frame dependent emissivity, $\alpha$ in the Bondi flow model. In the remaining two panels we plot $\ln\vartheta$ against the sGB coupling parameter $\zeta$ and the RZ deformation parameter $a_{1}$, for fixed powers of the emissivity. Solid lines are simulations with $\alpha = 6$, whereas dashed lines indicate $\alpha = 8$. Different colored lines represent different parameters of the flow (varying $\tilde{R}$ and $\gamma$) as shown in the legend.}
\label{fig:bonditheta}
\end{figure}

\subsection{Numerical experiments}

We now use the above results in our numerical experiments to better understand how $\vartheta$ changes with deformations to GR and other astrophysical parameters. As in the previous sections of the paper, we assume the plasma is optically thin, and further choose $\dot{M} = 1$ without loss of generality. We use $\tilde{R} = 10$ and $\tilde{R} \gg 10$. Since the plasma flow inside the critical radius is cold, the $\tilde{R} = 10$ scenario captures an accretion flow where there is a transition between hot and cold flow in the simulation domain, whereas the $\tilde{R} \gg 10$ case captures a cold flow for the entire numerical integration and is very close to the cold accretion flow considered earlier. The adiabatic index is varied between  $\gamma = 13/9$, appropriate for a plasma with nonrelativistic ions and relativistic electrons, and $\gamma = 5/3$, appropriate for a nonrelativistic ideal monatomic gas. 
\par 

We run simulations of the Bondi flow using each combination of $(\tilde{R}, \gamma)$ as outlined above.  We use a root finder to solve \eqref{eq:ufinal} for the radial plasma flow velocity $u^{r}(r)$, which is evaluated at each point along the null ray trajectory, once the radial trajectory has been determined by ray tracing. We then use $u_{\mu}u^{\mu} = -1$ to determine $u^{t}(r)$ at each point of a given ray. This, along with the null ray four velocity, enables us to find the photon frequency in the plasma frame, which along with \eqref{eq:jnu} allows us to use \eqref{eq:radtrans} to find the intensity profile needed to compute $\vartheta$. As in \textsection~\ref{sec:radin}, $\alpha$, $\zeta$, and $a_{1}$ are varied over their previously defined ranges. The bottom panel in Fig.~\ref{fig:images} shows three example synthetic images, one in each gravity theory, using Bondi accretion. This figure shows that in Bondi accretion the shadow size and overall brightness profile are still minimally changed by modifying the spacetime background.

\par 

The results of our simulations are also shown in Figure~\ref{fig:bonditheta}. It is clear from inspection that the pattern seen in \textsection~\ref{sec:radin} holds for the Bondi case; $\vartheta$ remains significantly more sensitive to the choice of astrophysical emission model than the spacetime metric. Furthermore, $\vartheta$ is more or less unaffected by the choice of accretion parameters, as the choice of $\tilde{R}$ and $\gamma$ make very little difference in the value of $\vartheta$ for a given $\alpha$, $\zeta$ or $a_{1}$. This simple calculation drives home the fact that $\vartheta$ is far more sensitive to the choice of astrophysical emission model than the deformation to the GR spacetime metric.

\section{Phenomenological Model} \label{sec:phenom}

In this section, we systematically study the intensity profiles in GR, sGB and the RZ spacetime in the radially infalling gas accretion model to understand why $\vartheta$ is largely unaffected by our choice of gravity theory. In \textsection~\ref{sub:analytic}, we derive a piecewise analytic function for the intensity profile $I(b)$ and use this intensity profile to compute $\vartheta$ by evaluating \eqref{eq:rchardef} numerically.
\par 
In \textsection~\ref{sub:bbc_comparison}, we show that recasting our numerical intensity profiles from $I(b) \to I(b/b_{c})$ results in striking overlap between intensity profiles in GR, sGB, and in the RZ metric. We confirm that this transformation can be done in any theory of gravity, and further prove the following statement: if the intensity profile $I(b)$ can be expressed as $I(b/b_{c})$, then $\vartheta$ is independent of the critical impact parameter $b_{c}$. This is why $\vartheta$ is nearly independent of spacetime metric in \textsection~\ref{sec:radin} and \textsection~\ref{sec:bondi}, and the only corrections to $\vartheta$ must arise from corrections to the functional form of $I(b/b_{c})$.

\subsection{Generic analytic calculation of intensity profile} \label{sub:analytic}

Here we use the cold accretion model, with $u_{\mu} = \left( -1, \sqrt{- g_{rr}(1 + g^{tt})}, 0, 0\right)$.  We compute the intensity profile in three distinct regimes: (I) the region interior to the critical impact parameter, (II) the region exterior to the critical impact parameter, and (III) the region asymptotically far away from the source. In both region I and region II, we compute the profile by approximating the integral 
\begin{equation}
    \mathcal{I} = \int_{\lambda_{0}}^{\lambda_{1}} J(\lambda) d\lambda, \label{eq:genintegral}
\end{equation}
using the Laplace approximation. We begin with interior to the critical curve.

\subsubsection{Region I: Interior to the critical curve} \label{subsub:inside}

\begin{figure}
    \centering
    \includegraphics[width=\columnwidth]{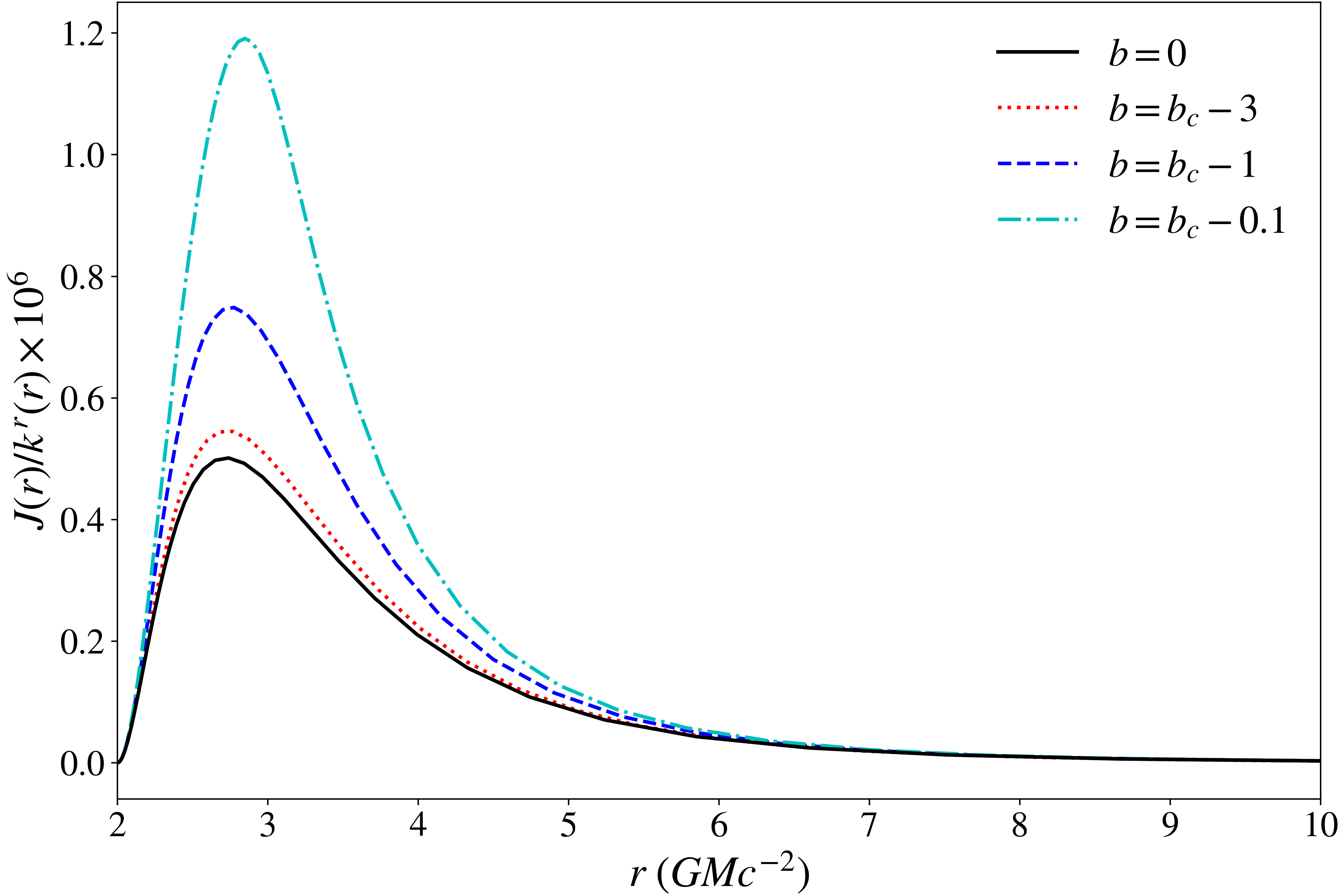}
    \caption{The integrand of \eqref{eq:genintegral_INT} plotted against the radial coordinate $r$ for numerous choices of $b < b_{c}$ in the GR radially infalling gas accretion scenario with $\alpha = 6$. Notice that the integrand is sharply peaked at approximately $r = r_{ph} = 3$ for each choice of \textit{b}, making Laplace's approximation suitable. This trend was checked to be true for all $\alpha$ considered in this work, as well as for sGB and RZ simulations with appropriate values for $r_{ph}$ in those spacetimes.}
    \label{fig:emiss_interior}
\end{figure}

Before beginning our approximation of \eqref{eq:genintegral} in this section, we transform \eqref{eq:genintegral} from an integral over the affine parameter, $\lambda$, to an integral over the radial coordinate, \textit{r}, by use of the fact that $k^r = dr/d\lambda$. Hence, in this section we aim to approximate the integral,
\begin{equation}
    \mathcal{I} = \int_{r_H}^{\infty} \dfrac{J}{k^r} dr, \label{eq:genintegral_INT}
\end{equation}
where the bounds have been chosen such that rays begin at the horizon and end at our telescope infinitely far away from the central black hole. Numerically evaluating the integrand of \eqref{eq:genintegral_INT} for numerous choices of $b < b_{c}$ reveals that for any \textit{b} the integrand is sharply peaked around $r = r_{ph}$; this is shown in Figure~\ref{fig:emiss_interior}. Hence, we approach approximating \eqref{eq:genintegral_INT} in the following way. We derive a generic expression for $J/k^r$ in terms of the radial coordinate only. Then we Taylor expand $J/k^r$ about $r = r_{ph}$; use of the zeroth order term and second order term in this expansion allows us to gain an approximation for \eqref{eq:genintegral_INT}, as desired.

\begin{figure*}
    \centering
    \includegraphics[width=\textwidth]{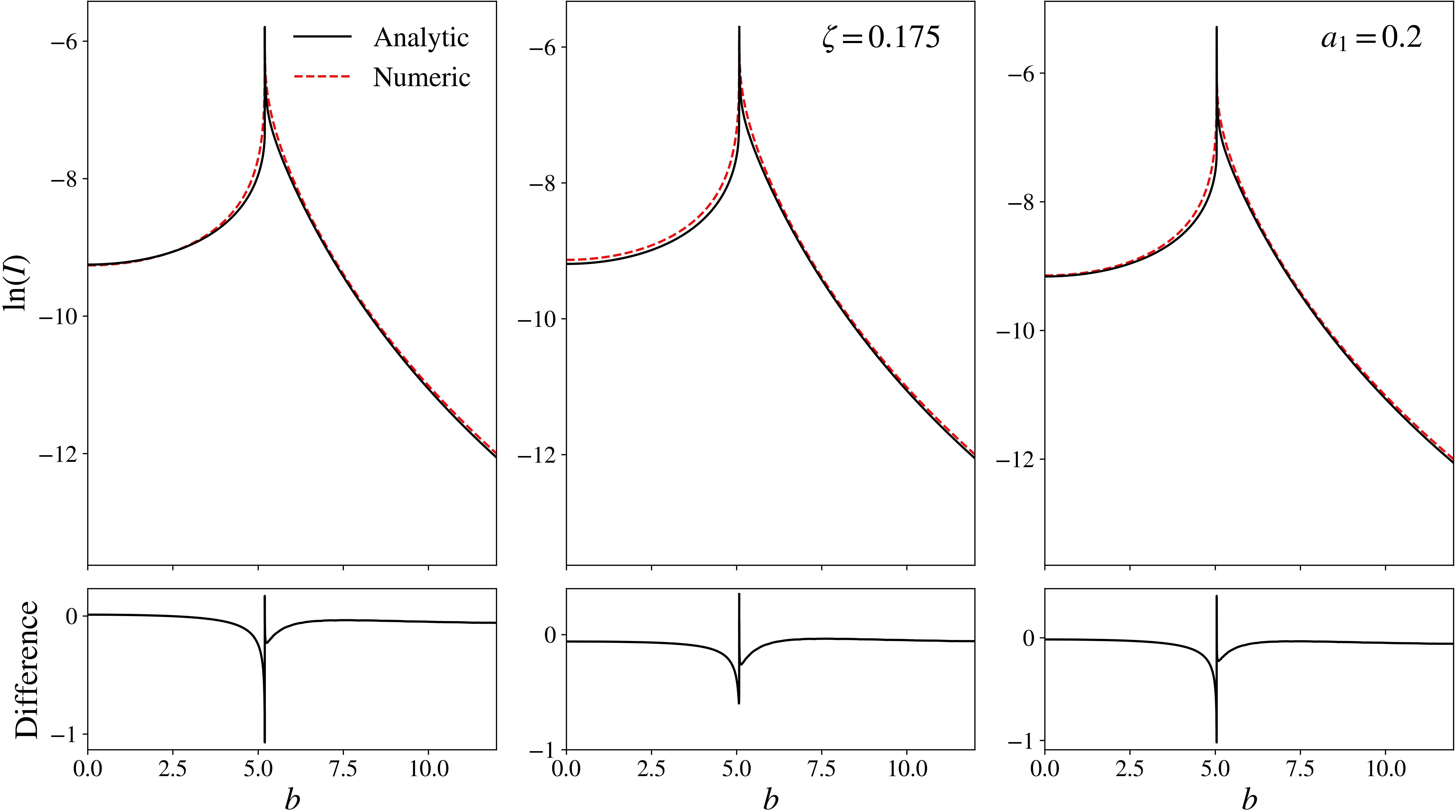}
    \caption{ (Top panels) The logarithmic intensity profile computed using \eqref{eq:interiorfinal} for $b < b_{c}$ and \eqref{eq:intensityclose} for $b > b_{c}$ in GR (left), sGB gravity (middle), and with the RZ metric (right), compared directly to our numerical result, for $\alpha = 6, \zeta = 0.175$ and $a_{1} = 0.2$. In the bottom panels we plot the relative difference between the analytic and numerical result. This comparison shows great agreement between full numerical simulations and analytic calculations.}
    \label{fig:num_ana_comp}
\end{figure*}

Using the definition of energy, $E = -g_{\mu\nu} k^{\mu}\xi^{\nu}_{(t)}$ and angular momentum, $L = g_{\mu\nu}k^{\mu}\xi^{\nu}_{(\phi)}$, where $\xi^{\mu}_{(t)}$ and $\xi^{\mu}_{(\phi)}$ are Killing vectors associated with time translation and rotation symmetry, respectively, we can write the photon four momentum as $k_{\mu} = (-E, k_{r}(\lambda), 0, b E)$. Note that we have used $b = L/E$ and taken $\theta = \pi/2$ without loss of generality. Using the four momentum normalization condition $k_\mu k^\mu = 0$, 
\begin{equation}
k_r(r) = - \dfrac{E}{r} \sqrt{- \dfrac{g_{rr} \ell(r,b)}{g_{tt}}} \label{eq:interior_wavevector}
\end{equation}
where
\begin{equation}
    \ell(r,b) := r^2 + b^2 g_{tt}.
\end{equation}
Raising the indices on the four momentum and using \eqref{eq:interior_wavevector}, $\nu = - k^{\mu} u_{\mu}$ becomes  
\begin{equation}
\nu = - \dfrac{E}{g_{tt}} \left( \dfrac{\sqrt{(1 + g_{tt})\ell(r,b) }}{r} - 1 \right). \label{eq:freq_interior}
\end{equation}
Recalling that $J := j_\nu / \nu^{2}$, we can use \eqref{eq:interior_wavevector}, \eqref{eq:freq_interior} and \eqref{eq:jnu} to write \eqref{eq:genintegral_INT} as 
\begin{equation}
    \mathcal{I} = \dfrac{j_0 r_0^\alpha}{E^3} \int_{r_H}^{\infty} \dfrac{r^{3-\alpha} g_{rr} g_{tt}^{5/2} dr}{\sqrt{-g_{rr} \ell(r,b)} \left( r - \sqrt{(1+g_{tt})\ell(r,b)})\right)^{2}} \label{eq:invarintens_int}
\end{equation}
after simplification. Using $\mathcal{I} = I/\nu^{3}$ and that $\nu = E$ at infinity, \eqref{eq:invarintens_int} becomes
\begin{equation}
         I(\alpha, b) = \int_{r_H}^{\infty} \mathcal{A}(r,b,\alpha) dr
\end{equation}
where
\begin{equation}
    \mathcal{A}(r,b,\alpha) := \dfrac{j_0 r_0^\alpha r^{3-\alpha} g_{rr} g_{tt}^{5/2}}{\sqrt{-g_{rr} \ell(r,b)} \left( r - \sqrt{(1+g_{tt})\ell(r,b)})\right)^{2}}. \label{eq:A}
\end{equation}
We now divide out the slowly varying components of \eqref{eq:A} before applying the Laplace approximation. In general, this step is equivalent to dividing \eqref{eq:A} by $r^{\eta}$, where $\eta > 0 $; in GR, $\eta = 7/2 - \alpha$. We keep $\eta$ as an undetermined parameter for the remainder of the derivation, but used $\eta = 7/2 - \alpha$ for the profiles found in Appendix~\ref{app:interior}. Defining
\begin{equation}
    \tilde{\mathcal{A}}(r,b,\alpha) := \ln\left( \dfrac{\mathcal{A}(r,b,\alpha)}{r^{\eta}} \right), \label{eq:Atilde}
\end{equation}
the Laplace approximation then gives
\begin{equation}
    I_{int}(\alpha, b) = \mathcal{A}(r_{ph}, b, \alpha) \sqrt{-\dfrac{2 \pi}{\partial_{r}^{2} \tilde{\mathcal{A}}(r,b,\alpha)\vert_{r=r_{ph}}}}. \label{eq:interiorfinal}
\end{equation}
This completes the derivation for the interior of the critical curve. In Appendix~\ref{app:interior}, we evaluate \eqref{eq:interiorfinal} in GR, and to first order in the coupling constants in sGB and RZ. This calculation agrees with our numerical results to striking accuracy, see Figure~\ref{fig:num_ana_comp}.

\subsubsection{Region II: Outside the critical curve} \label{subsub:crit}

\begin{figure}
    \centering
    \includegraphics[width=\columnwidth]{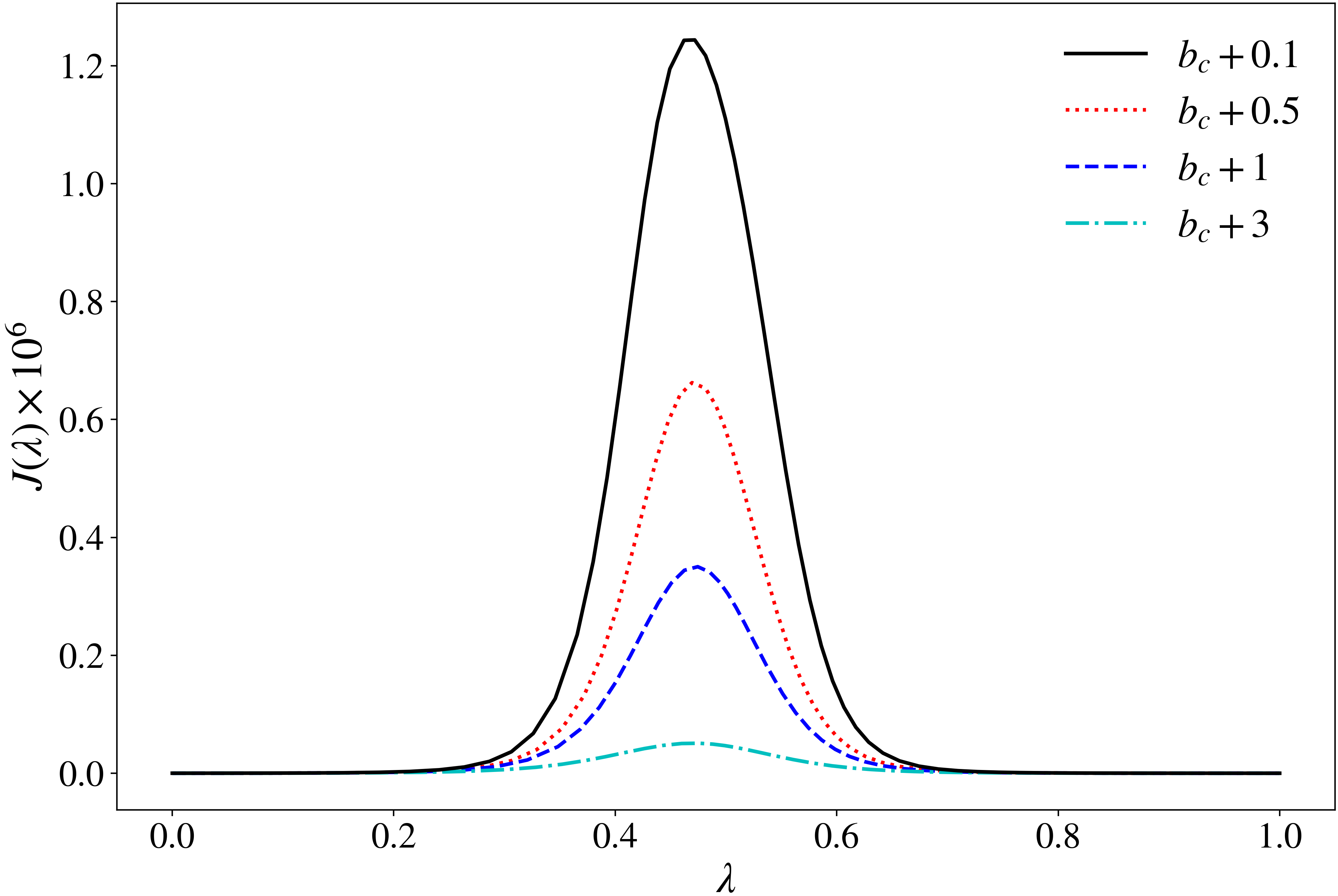}
    \caption{The invariant emissivity, $J(\lambda)$, plotted against the affine parameter $\lambda$ for numerous choices of $b > b_{c}$ in the GR radially infalling gas accretion scenario with $\alpha = 6$. Notice that $J(\lambda)$ is sharply peaked for each choice of \textit{b}, making Laplace's approximation suitable. This trend was checked to be true for all $\alpha$ considered in this work, as well as for sGB and RZ simulations.}
    \label{fig:invaremiss}
\end{figure}

In the region near the critical curve (i.e., near the critical impact parameter), we derive the intensity profile generically using the following strategy. We evaluate the integral \eqref{eq:genintegral} by forming an expansion of the frame invariant emissivity, $J(\lambda)$, around the affine parameter which corresponds to the radius of closest approach, $\lambda_{min}$, such that $r(\lambda_{min}) = r_{min}$. Then, by exploiting the fact that $J(\lambda)$ can be well approximated by a Gaussian curve peaked at a point of maximum emission, (see Figure~\ref{fig:invaremiss} for different choices of $b>b_c$), we integrate \eqref{eq:genintegral} analytically using Laplace's method.

We begin our derivation by writing expansions about $\lambda_{min}$ for the radial profile $r(\lambda)$ and the photon four momentum $k_{\mu}(\lambda)$. Note that for each null ray outside of the photon sphere there exists a unique radius of closest approach, $r_{min}$. Furthermore, since each spacetime we consider is spherically symmetric, the path taken from $r \gg r_{min}$ to $r_{min}$ must be identical to that taken from $r_{min}$ to $r \gg r_{min}$. This implies that $r(\lambda)$ is an even function of the affine parameter about $\lambda_{min}$. Allowing $r(\lambda_{min}) = r_{min}$, it follows that, to lowest order in $(\lambda - \lambda_{min})$ we can approximate $r(\lambda)$ as 
\begin{equation}
    r(\lambda) \approx r_{min} + \dfrac{(\lambda - \lambda_{min})^{2}}{2} \dfrac{d^{2}r(\lambda)}{d\lambda^{2}} \bigg\vert_{\lambda_{min}}. \label{eq:ransatz}
\end{equation}
Again exploiting time and azimuthal translation symmetries allows us to write $k_{\mu} = (-E, k_{r}(\lambda), 0, b E)$. It immediately follows from \eqref{eq:ransatz} that the radial component of the photon four momentum is linear in $(\lambda - \lambda_{min})$, i.e.,
\begin{equation}
    k_{r}(\lambda) \approx (\lambda - \lambda_{min}) \dfrac{dk_{r}(\lambda)}{d\lambda}\bigg\vert_{\lambda_{min}}. \label{eq:kransatz}
\end{equation}
Note that, given \eqref{eq:ransatz} and \eqref{eq:kransatz}, the limit $\lambda \to \lambda_{min}$ is equivalent to taking $r \to r_{min}$ and $k_{r} \to 0$. 

\par 

To solve for $d^{2}r(\lambda)/d\lambda^{2}$ and $dk_{r}/d\lambda$ at $\lambda = \lambda_{min}$, we use the null condition, $k_{\mu}k^{\mu} = 0$. Expanding the null condition, we find that 
\begin{equation}
    \dfrac{b^{2}E^{2}}{r^{2}} + \dfrac{k_{r}^{2}}{g_{rr}} + \dfrac{E^{2}}{g_{tt}}  = 0. \label{eq:nullcondition}
\end{equation}
Evaluating \eqref{eq:nullcondition} in the limit of $\lambda \to \lambda_{min}$, we recover 
\begin{equation}
    b^{2} = - \dfrac{r^{2}}{g_{tt}} \bigg\vert_{\lambda_{min}}. \label{eq:brelate}
\end{equation}
We now use \eqref{eq:nullcondition} to solve for $dk_{r}(\lambda)/d\lambda$ at $\lambda = \lambda_{min}$ by differentiating \eqref{eq:nullcondition} with respect to the affine parameter and evaluating at $\lambda = \lambda_{min}$, which results in
\begin{equation}
    \dfrac{dk_{r}(\lambda)}{d\lambda}\bigg\vert_{\lambda_{min}} = \dfrac{E^{2} \mathcal{R}(r)}{2r g_{tt}^{2}} \bigg\vert_{\lambda_{min}}, \label{eq:krpmin}
\end{equation}
where we have used \eqref{eq:brelate} to simplify \eqref{eq:krpmin} and defined 
\begin{equation}
    \mathcal{R}(r) := r \dfrac{\partial g_{tt}}{\partial r} - 2 g_{tt}. \label{eq:ryansfunction}
\end{equation}
We now raise the index of \eqref{eq:krpmin} to find 
\begin{equation}
    \dfrac{d^{2}r(\lambda)}{d\lambda^{2}}\bigg\vert_{\lambda_{min}} = \dfrac{E^{2} \mathcal{R}(r)}{2 r g_{rr} g_{tt}^{2}}\bigg\vert_{\lambda_{min}}. \label{eq:rppmin}
\end{equation}
Equations~\eqref{eq:krpmin} and \eqref{eq:rppmin} allow us to evaluate the frame dependent emissivity \eqref{eq:jnu} and the frequency, $\nu = - k_{\mu}u^{\mu}$. Using \eqref{eq:rppmin} in \eqref{eq:jnu} and expanding around $\lambda_{min}$, we have 
\begin{align}
    j_{\nu} \approx j_{0} \left( \dfrac{r_{0}}{r_{min}}\right)^{\alpha} \bigg( 1 - \dfrac{\alpha (\lambda - \lambda_{min})^{2}}{2 r_{min}} \dfrac{d^{2}r(\lambda)}{d\lambda^{2}}\bigg\vert_{\lambda_{min}}\bigg). \label{eq:framedepemiss_sub}
\end{align}
Using \eqref{eq:krpmin} we find that, to lowest order in $(\lambda - \lambda_{min})$, the frequency is given by
\begin{equation}
    \nu \approx \dfrac{E}{g_{tt}}\bigg\vert_{\lambda_{min}} + \nu^{(1)}\bigg\vert_{\lambda_{min}} (\lambda - \lambda_{min}), \label{eq:nu}
\end{equation}
where 
\begin{equation}
    \nu^{(1)}(r) := \sqrt{- \left(1 + \dfrac{1}{g_{tt}}\right)} \dfrac{dk_{r}(\lambda)}{d\lambda}. \label{eq:nu1}
\end{equation}
Note that in \eqref{eq:nu} we have neglected second order contributions to the frequency because the frame invariant emissivity, $J$, goes as $J \sim \nu^{-2}$, making any $\mathcal{O}[(\lambda - \lambda_{min})^{2}]$ terms too high order to contribute to the final calculation. 

\par 

We now use \eqref{eq:framedepemiss_sub} and \eqref{eq:nu} in \eqref{eq:radtrans}, to find that the invariant emissivity can be expanded to second order in $(\lambda - \lambda_{min})$ as
\begin{align}
    J(\lambda) & \approx j_{0} \left( \dfrac{r_{0}}{r_{min}} \right)^{\alpha} g_{tt}^{2}\bigg\vert_{\lambda_{min}} \bigg( \dfrac{1}{E^{2}} \nonumber \\
    & + J^{(1)}\bigg\vert_{\lambda_{min}} (\lambda - \lambda_{min}) \nonumber \\ 
    & + J^{(2)}\bigg\vert_{\lambda_{min}} \dfrac{(\lambda - \lambda_{min})^{2}}{2} \bigg), \label{eq:invaremis}
\end{align}
where we have defined
\begin{align}
    J^{(1)}(r) & := - \dfrac{\sqrt{-g_{rr} (1 + g^{tt})} \mathcal{R}(r)}{r g_{rr} g_{tt} E}, \label{eq:j1} \\
    J^{(2)}(r) & := \dfrac{\mathcal{R}(r) \mathcal{N}(r)}{2 g_{rr} r^{2} g_{tt}^{3}},\label{eq:j2}
\end{align}
and
\begin{equation}
    \mathcal{N}(r) := 6 g_{tt}^{2} - (\alpha - 6) g_{tt} - 3 (1 + g_{tt}) r \dfrac{\partial g_{tt}}{\partial r}.\label{eq:Ndefn}
\end{equation}

With \eqref{eq:invaremis} in hand, we are now able to apply the Laplacian approximation. To calculate the point of maximum emission, we calculate the roots of the $dJ(\lambda)/d\lambda$ from \eqref{eq:invaremis}, resulting in
\begin{equation}
\lambda_{max} = \lambda_{min}-\dfrac{J^{(1)}}{J^{(2)}}\bigg\vert_{\lambda_{min}} ,
\end{equation} 
which means the expansion of the invariant emissivity about the point of maximum emission, $\tilde{J}(\lambda)$ is given by
\begin{align}
    \tilde{J}(\lambda) & \approx j_{0} \left( \dfrac{r_{0}}{r_{min}}\right)^{\alpha}g_{tt}^{2}\bigg\vert_{\lambda_{min}}\Bigg( \dfrac{J(\lambda_{max})}{E^{2}} \nonumber \\ 
    & + \dfrac{(\lambda - \lambda_{max})^{2}}{2} J^{(2)}\bigg\vert_{\lambda_{min}} \Bigg),
\end{align}
where 
\begin{align}
    J(\lambda_{max}) & := \dfrac{1}{E^{2}} \left( 1 + \dfrac{(1+g^{tt}) \mathcal{R}(r)}{\mathcal{N}(r)}\right). \label{eq:joffset}
\end{align}
The invariant intensity is then given by
\begin{equation}
    \mathcal{I}(r_{min}) = \dfrac{J(\lambda_{max})}{E^{3}} \sqrt{ - \dfrac{ 2\pi J(\lambda_{max})}{ J^{(2)}|_{\lambda_{min}}}}. \label{eq:invarintensity}
\end{equation}
Evaluating \eqref{eq:invarintensity} in the telescope frame, we arrive at

\begin{equation}
    I_{ext}(r_{min}) =  J(\lambda_{max}) \sqrt{-\dfrac{2 \pi J(\lambda_{max})}{ J^{(2)}|_{\lambda_{min}}}}, \label{eq:intensityclose}
\end{equation}
as desired. This concludes this component of the calculation. A functional form of the intensity profile as a function of the impact parameter, \textit{b} can be obtained by noting that \eqref{eq:brelate}, for all $b > b_{c}$ (i.e., $r_{min} > r_{ph}$), provides a one-to-one mapping between impact parameter and radius of closest approach. Therefore, we choose to leave our result in terms of $r_{min}$ for simplicity. In Appendix~\ref{app:intensity}, we evaluate \eqref{eq:intensityclose} in GR, sGB, and in the RZ metric for reference. Notice that the GR intensity profile diverges as $\left(r_{min} - 3 \right)^{-1/2}$ near the photon sphere (i.e., for $b \approx b_{c}$), whereas the sGB and RZ first-order corrections diverge as $\left( r_{min} - 3 \right)^{-3/2}$.

\subsubsection{Region III: Asymptotically far away} \label{subsub:asymp}

We now consider the last regime of the calculation: asymptotically far away from the black hole. In this region, null rays travel on straight lines; without loss of generality, assume photons begin at $z \to - \infty$ and arrive at our telescope at $z \to \infty$. We invoke a cylindrical coordinate system, where the impact parameter of a null ray $b$ plays the roll of the cylindrical radius in the image. In this framework, the spacetime is described by Minkowski space, making $k^{t} = E$ by the definition of energy, and $k^{z} = E$ by the null condition. As noted in \textsection~\ref{subsub:inside}, the frequency asymptotically far away is $\nu = E$. Furthermore, the radial coordinate in spherical coordinates is given by $r = \sqrt{b^{2} + z^{2}}$ in our cylindrical coordinate system, making \eqref{eq:radtrans} in this regime
\begin{equation}
\mathcal{I} = \dfrac{j_{0}r_{0}^{\alpha}}{ E^{3}} \int_{-\infty}^{\infty} \dfrac{dz}{\left( b^{2} + z^{2}\right)^{\alpha/2}}. \label{eq:asympalmost} 
\end{equation}
Evaluating \eqref{eq:asympalmost} in the frame of our telescope and carrying out the integral, we find that
\begin{equation}
    I(b) = \dfrac{\Gamma\left( \dfrac{\alpha - 1}{2}\right)}{\Gamma\left( \dfrac{\alpha}{2} \right)} \dfrac{j_{0} r_{0}^{\alpha}\sqrt{\pi}}{b^{\alpha - 1}}, \label{eq:asympintensity}
\end{equation}
where $\Gamma(z)$ is the Gamma function. This completes the calculation of the intensity profile in region III. 

\subsubsection{Summary of Phenomenological Model and Validation}

For all $b < b_{c}$, the intensity is given by \eqref{eq:interiorfinal}; for $b > b_{c}$, the intensity is given by \eqref{eq:intensityclose}, after using \eqref{eq:brelate} to convert \eqref{eq:intensityclose} to a function over the impact parameter, \textit{b}. Finally, for $b \gg b_{c}$, one uses \eqref{eq:asympintensity} for the intensity. Figure~\ref{fig:num_ana_comp} shows the intensity profile computed analytically and numerically in GR, sGB, and using the RZ metric. The results are in great agreement for both the $b < b_{c}$ and $b > b_{c}$ regimes shown.

\par 

We can now use this phenomenological analytic profile to evaluate \eqref{eq:rchardef} numerically in GR, which allows us to compare calculations of $\vartheta$ done using both analytic and numerical methods. Regressing the logarithm of $\vartheta$ against $5 < \alpha \leq 8$, we find that 
\begin{equation}
    \vartheta_{GR} \sim \exp\left( -0.2 \alpha \right),
\end{equation}
which scales to the same order of magnitude as our numerical results in \textsection~\ref{sec:radin}, further validating our phenomenological model. 

\subsection{Intensity profile analysis in GR, sGB, and RZ} \label{sub:bbc_comparison}

In this section, we investigate the qualitative similarities between the intensity profiles in GR, sGB, and RZ. Our motivation for doing so is the following. Equations~\eqref{eq:bcritsGB} and \eqref{eq:bcritRZ} reveal that the critical impact parameter $b_{c}$ is linearly corrected in sGB and RZ by the deformation parameter multiplied by a constant of order unity. In \textsection~\ref{sec:radin} and \textsection~\ref{sec:bondi}, however, we found that $\vartheta$ is corrected by the deformation parameter multiplied by a coefficient that is $\sim 10^{-2}$. This implies that how $\vartheta$ deviates from its GR value does not only depend on the correction to $b_{c}$, but also the intensity profile that is integrated to find the characteristic radius of emission. Furthermore, there must be cancellation between the correction in $b_{c}$ and the correction to $r_{char}$. 

\par 

\begin{figure}
    \centering
    \includegraphics[width=\columnwidth]{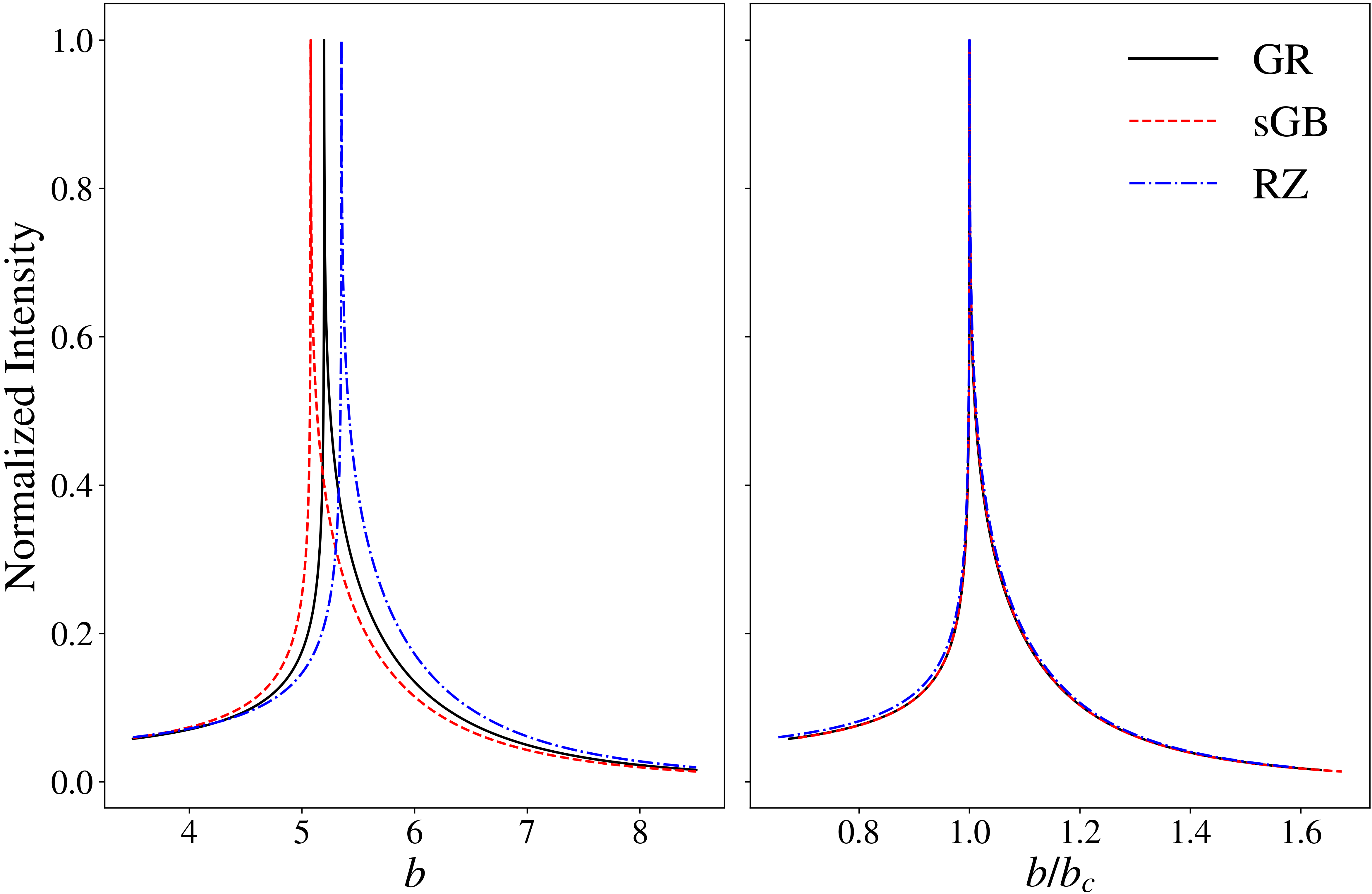}
    \caption{\textit{(Left).} Normalized intensity profiles in GR, sGB, and RZ plotted against the impact parameter \textit{b} in the radially infalling gas model with $\alpha = 6$. \textit{(Right).} We plot the same intensity profiles, but plotted against $b/b_{c}$, where $b_{c} = \sqrt{27}$ for GR, and $b_{c}$ was found by evaluating \eqref{eq:bcritsGB} with $\zeta = 0.175$ for sGB and \eqref{eq:bcritRZ} with $a_{1} = -0.2$ for RZ. Clearly, the intensity profile as a function of $b/b_{c}$ is near-identical for each theory of gravity }
    \label{fig:shiftedintensityprofiles}
\end{figure}

We propose that the reason for this cancellation lies in the covariance between $r_{char}$ and $b_{c}$. Since both sGB and RZ are continuous deformations to the Schwarzschild (GR) metric, the intensity profile (for a given accretion and emissivity model) has approximately the same magnitude as the GR intensity profile; i.e., sGB or RZ black holes do not ``glow" orders of magnitude brighter than GR black holes. The only qualitative difference between the profiles in GR, sGB, and RZ, therefore, is the location of their features; just as in GR, the sGB and RZ intensity profiles possess a peak near the critical impact parameter due to the increased path length of photons that orbit the black hole before reaching our detectors. Each profile also grows at approximately the same rate for $b < b_{c}$, and decays rapidly for $b \gg b_{c}$. Due to the relatively large amount of intensity generated near the critical impact parameter, and the fact that the characteristic radius of emission is defined as an integral over all impact parameters, it stands to reason that the characteristic radius of emission is dominated by the contribution of all impact parameters less than $b_{c}$ and just outside of $b_{c}$. 

\par 

Therefore, the characteristic radius of emission is only modified by changing the spacetime metric insofar as the critical impact parameter is modified, hence the insensitivity of $\vartheta$ on the choice of gravity model. 

\par 

In Figure~\ref{fig:shiftedintensityprofiles}, we illustrate  this by plotting the intensity profiles in GR, sGB and RZ against both the impact parameter \textit{b} and the ratio of the impact parameter to critical impact parameter $b/b_{c}$, where $b_{c}$ is adjusted to its appropriate value for each spacetime. We find striking overlap between the intensity profiles when plotted as functions of $b/b_{c}$, which supports our claim that the main difference between intensity profiles is the location of their features, with the functional form of the intensity remaining approximately the same. This drives home that the characteristic radius of emission and the critical impact parameter must covary, causing $\vartheta$ to change significantly less than the critical impact parameter when the spacetime metric is modified. 

\par 

We can also support our claim through an analytical exercise. Suppose we model our intensity profile as a function of $b/b_{c}$. Using our results in \textsection~\ref{subsub:inside}, we see that this transformation is straightforward in the regime of $b < b_{c}$, and in \textsection~\ref{subsub:crit} we proved this can be done in the regime of $ b > b_{c}$ through \eqref{eq:brelate} after dividing both sides by $b_{c}^{2}$. In both of the cases mentioned above the functional form of the metric functions in the static, spherically symmetric spacetime were kept general, and hence this transformation can be done for any such spacetime. By the definition of the characteristic radius of emission, \eqref{eq:rchardef}, we then have
\begin{equation}
    \vartheta = \sqrt{\dfrac{\int_{0}^{\infty} \xi^{3} I(\xi) d\xi}{\int_{0}^{\infty} \xi I(\xi) d\xi}}, \label{eq:varthetaindep}
\end{equation}
where we have changed variables to $\xi := b/b_{c}$. This explicitly proves that $\vartheta$ is independent of $b_{c}$ entirely. Furthermore, \eqref{eq:varthetaindep} implies that any corrections from the spacetime metric do not enter through the critical impact parameter, but rather, through the functional form of the intensity profile $I(\xi)$, which captures the astrophysical uncertainties through $\alpha$ and the GR modifications via its dependence on the metric functions.

\par 

Let us now consider a generic deformation to the $(t,t)$ and $(r,r)$ components of a spherically symmetric spacetime metric, the strength of which is mediated by a non-dimensional coupling constant $\varepsilon$, i.e.~$g_{tt} = g_{tt}^{GR} + \varepsilon \tilde{g}_{tt} + {\cal{O}}(\varepsilon^2)$ and $g_{rr} = g_{rr}^{GR} + \varepsilon \tilde{g}_{rr} + {\cal{O}}(\varepsilon^2)$. Using these metric functions in \eqref{eq:interiorfinal} and \eqref{eq:intensityclose} will result in an intensity profile $I(\xi)$ of the form $I(\xi) = I_{GR}(\xi) + \varepsilon \tilde{I}(\xi) + {\cal{O}}(\varepsilon^2)$ after expansion to first order in $\varepsilon$. Using this ansatz in \eqref{eq:varthetaindep}, 
\begin{equation}
    \vartheta = \sqrt{\dfrac{M_{3}(I_{GR}) + \varepsilon M_{3}(\tilde{I})}{M_{1}(I_{GR}) + \varepsilon M_{1}(\tilde{I})}}, \label{eq:vartheta_moments}
\end{equation}
where we have introduced the shorthand notation of 
\begin{equation}
    M_{n}(f) := \int_{0}^{\infty} x^{n} f(x) dx,  \label{eq:momentdefn}
\end{equation}
for any function \textit{f}. One can think of $M_n(I)$ as the $n$th moment of the intensity function. Expanding \eqref{eq:vartheta_moments} to first order in $\varepsilon$, we find that 
\begin{equation}
    \vartheta = \vartheta_{GR} \left( 1 + \dfrac{\varepsilon}{2} \left( \dfrac{M_{3}(\tilde{I})}{M_{3}(I_{GR})} - \dfrac{M_{1}(\tilde{I})}{M_{1}(I_{GR})} \right) \right),
\end{equation}
where
\begin{equation}
    \vartheta_{GR} := \sqrt{\dfrac{M_{3}(I_{GR})}{M_{1}(I_{GR})}}. \label{eq:varthetagr}
\end{equation}
We can calculate this quantity explicitly using our phenomenological model, taking \eqref{eq:GRINT} as the GR intensity profile for $b < b_c$ and \eqref{eq:grprofile} for $b > b_c$; we find that, for $\alpha = 6$, $M_{3}(I_{GR}) \approx 5.20 \times 10^{-4}$ and  $M_{1}(I_{GR}) \approx 2.99 \times 10^{-4}$, and therefore $\vartheta_{GR} \approx 1.32$, which differs from our numerical result by $\sim$12\%. We can therefore explore corrections to $\vartheta$ induced by a deformation of our metric components by examining the quantity 
\begin{equation}
    \delta \vartheta(\tilde{I}, I_{GR}) := \dfrac{1}{2} \left(\dfrac{M_{3}(\tilde{I})}{M_{3}(I_{GR})} - \dfrac{M_{1}(\tilde{I})}{M_{1}(I_{GR})} \right), \label{eq:deltvartheta}
\end{equation}
such that
\begin{equation}
    \vartheta = \vartheta_{GR} (1 + \delta\vartheta(\tilde{I},I_{GR}) \varepsilon).
\end{equation}

We can evaluate \eqref{eq:deltvartheta} in sGB gravity and using the RZ metric in the following way. First, choose an $\alpha > 5$; we choose $\alpha = 6$, for direct comparison with our numerical results in \textsection~\ref{sec:radin}. We can evaluate $M_{n}(\tilde{I})$ over the three distinct regimes defined in \textsection~\ref{sec:phenom}: inside the critical impact parameter, outside the critical impact parameter, and asymptotically far away from the black hole. However, asymptotically far away from the black hole, $\tilde{I} = 0$, as we assumed our spacetimes limit to Minkowski (flat space) in the asymptotic regime, therefore resulting in no corrections due to GR modifications in this limit. Hence, it is only necessary to evaluate $M_{n}(\tilde{I})$ in the first two regions of our analytic calculation; the sum of the moment in each region gives us the total moment for the intensity profile in consideration.
\par 
We begin interior to the critical impact parameter, i.e., for $0 < \xi < 1$. In this region, we approximate the intensity as given by \eqref{eq:interiorfinal}. We use \eqref{eq:SGBCorInt} as the correction to the intensity profile in sGB and \eqref{eq:RZCorInt} as the correction in RZ to find that $M_{3}(\tilde{I}_{sGB}^{int}) \approx 2.00 \times 10^{-4}$, and $M_{1}(\tilde{I}_{sGB}^{int}) \approx 1.53 \times 10^{-4}$, whereas in RZ we have $M_{3}(\tilde{I}_{sGB}^{int}) \approx 9.80 \times 10^{-5}$, and $M_{1}(\tilde{I}_{sGB}^{int}) \approx 1.26 \times 10^{-4}$. 

\par 

For $\xi > 1$, we use \eqref{eq:mathcale} as the profile correction for sGB and \eqref{eq:rzcorrection} as the correction using the RZ metric. Note that this integration is equivalent to integrating the intensity profile in the domain of $r_{min} > r_{ph}$. Far away from the black hole (i.e., when $r_{min}$ is large), it can be shown that the GR intensity profile given in \eqref{eq:grprofile} decays as $\sim r_{min}^{-(\alpha - 1)}$, whereas the correction to the sGB profile \eqref{eq:mathcale} decays as $\sim r_{min}^{-(\alpha + 1)}$ and the RZ correction \eqref{eq:rzcorrection} decays as $\sim r_{min}^{-(\alpha + 2)}$. 
\par 
Two insights can be gleaned from this analysis. First, the GR intensity profile decays at the same rate as the Minkowski profile found in \eqref{eq:asympintensity}. Secondly, the corrections to the analytic intensity profile due to sGB and RZ deformations are of lower order than the GR intensity profile at large $r_{min}$. Therefore, in practice, the corrections to the GR intensity profile decay quickly, and the profile given by \eqref{eq:grprofile} dovetails with \eqref{eq:asympintensity}, allowing us to use the intensity profile in region II for $ \xi > 1$ all the way out to $r_{min} \to \infty$, never needing to use \eqref{eq:asympintensity} in our calculations.

With this in mind, we can now evaluate the moments in the middle region. For the sGB metric, we find that $M_3(\tilde{I}_{sGB}^{ext}) \approx 8.30 \times 10^{-5}$ and $M_1(\tilde{I}_{sGB}^{ext}) \approx 8.07 \times 10^{-5}$, while for the RZ metric we find $M_3(\tilde{I}_{RZ}^{ext}) \approx - 2.99 \times 10^{-6}$ and $M_1(\tilde{I}_{RZ}^{ext}) \approx -8.24 \times 10^{-6}$. The numerical values of $M_3$ and $M_1$ in GR, sGB and in the RZ spacetime are succinctly presented in Table~\ref{tab:moments}.
\par 
From these calculations we glean that it is not that the two terms responsible for $\delta\vartheta$ (i.e.~the GR-normalized moments) nearly cancel each other, but rather that each of them is separately small. Indeed, $M_3(\tilde{I})/M_3(I_{GR}) \sim {\cal{O}}(10^{-1}) \sim M_1(\tilde{I})/M_1(I_{GR})$ for the sGB metric, while $M_3(\tilde{I})/M_3(I_{GR}) \sim {\cal{O}}(10^{-1})$ and $ M_1(\tilde{I})/M_1(I_{GR}) \sim \mathcal{O}(10^{-2})$ for the RZ metric. The fact that one must take the difference between these two normalized moments further shrinks $\delta \vartheta$, but it is their intrinsic smallness that explains why $\delta \vartheta \ll 1$. In turn, the reason for this smallness is because in the regime $b > b_c$, $M_n(\tilde{I}) < M_n(I_{GR})$, and this is because the corrections to the intensity profile in the modified metrics decay faster than the profile of GR for large $r_{min}$, as previously discussed. Indeed, as one can clearly see from Table~\ref{tab:moments}, the moment in region II dominates the GR moment calculation, but the moment contribution from region II is of the same magnitude or lower than region I for the sGB and RZ calculations hence a smaller overall $M_{n}(\tilde{I})$ in both cases. This claim further is supported by the fact that the decay rate of $\tilde{I}_{sGB}$ is slower than $\tilde{I}_{RZ}$ in region II, and $M_{3}(\tilde{I}_{sGB}), M_{1}(\tilde{I}_{sGB}) \sim \mathcal{O}(10^{-5})$ in this region, whereas $M_{3}(\tilde{I}_{RZ}), M_{1}(\tilde{I}_{RZ}) \sim \mathcal{O}(10^{-6})$.

\par 

By evaluating \eqref{eq:deltvartheta} using the results in Table \ref{tab:moments}, we recover the following for $\vartheta$ in sGB and using the RZ metric,
\begin{align}
    \vartheta_{sGB} & = \vartheta_{GR} \left( 1 + 0.13 \zeta \right), \\
    \vartheta_{RZ} & = \vartheta_{GR} \left( 1 - 0.10 a_{1} \right),
\end{align}
where $\vartheta_{GR} \approx 1.32$ is the GR value of $\vartheta$ when $\alpha = 6$. This coincides with our previous predictions: the correction to $\vartheta$ induced by a modification to gravity is $\sim \mathcal{O}(10^{-2})$ or lower, since $\zeta,a_{1} \sim \mathcal{O}(10^{-1})$.

\begin{table*}[!htbp]
    \centering
    \begin{tabular}{|c||c|c|c||c|c|c||c|c|c|}
    \hline
         {} & \multicolumn{3}{c||}{GR} & \multicolumn{3}{c||}{sGB} & \multicolumn{3}{c|}{RZ} \\
         \hline \hline 
         {} & I & II & Total & I & II & Total & I & II & Total \\
         \hline  
         $M_{1}$ & 9.32 $\times 10^{-5}$ & 2.05 $\times 10^{-4}$ & 2.99 $\times 10^{-4}$ & 1.53 $\times 10^{-4}$ & 8.07 $\times 10^{-5}$ & 2.80 $\times 10^{-4}$ & 1.26 $\times 10^{-4}$ & -8.24 $\times 10^{-6}$ & 1.17 $\times 10^{-4}$ \\
         $M_{3}$ & 5.86 $\times 10^{-5}$ & 4.62 $\times 10^{-4}$ &  5.20 $\times 10^{-4}$ & 2.00 $\times 10^{-4}$ & 8.30 $\times 10^{-5}$ & 2.36 $\times 10^{-4}$ & 9.80 $\times 10^{-5}$ & -2.99 $\times 10^{-6}$ & 9.50 $\times 10^{-5}$ \\
         \hline  
    \end{tabular}
    \caption{The numerical values of the first and third moment of the intensity profile in GR, sGB and RZ, evaluated in region I (interior to the critical curve) and region II (exterior to the critical curve). The total moment is also given, which is the sum of the moment in region I and region II. Note that all values in this table were computed for $\alpha = 6$.}
    \label{tab:moments}
\end{table*}

\section{Discussion} \label{sec:discuss}

In this paper, we calculated the ratio of the characteristic radius of emission and the critical impact parameter, defined as $\vartheta$, to investigate the sensitivity of EHT observables to variations in the astrophysical emission model, the gravity theory, and the properties of the accreting plasma. To probe the sensitivity of $\vartheta$ to the choice of gravity theory, we compute $\vartheta$ in GR, sGB gravity and the RZ metric. We conducted this analysis using two accretion models: radial infall and Bondi flow. The findings of our study show that $\vartheta$ is primarily affected by the choice of emission model. Indeed, the impact of variations in the frame dependent emissivity trump the corrections induced by the modified gravity theory and minor variations in the plasma inflow model. 
\par 
Our interpretation of these results is the following. Since the intensity profile is peaked near the critical impact parameter, most of the contribution to the calculation of the characteristic radius of emission comes from near the critical impact parameter. This implies that the location of the characteristic radius of emission only changes insofar as the critical impact parameter changes due to the modified gravity theory. Therefore, it is challenging to resolve a deviation from GR, given the covariance of the characteristic radius of emission and the critical impact parameter. This explains why $\vartheta$ is weakly impacted by departures from GR.
\par 
We further supported this conclusion by finding a piecewise analytic approximation for the intensity profile $I(b)$ in radially infalling gas accretion model. We showed that our intensity profile matches the numerical results accurately in each theory of gravity considered in this work. We then showed the intensity profile can always be cast in the form $I(b/b_{c})$, and proved that if the intensity profile $I(b)$ can be expressed as a function of the ratio between the impact parameter and critical impact parameter $b/b_{c}$, then $\vartheta$ is independent of $b_{c}$. Therefore, any corrections to $\vartheta$ must come from the functional form of $I(b/b_{c})$, which depends on the metric functions rather than $b_{c}$. 
\par

This result allows us to weigh in on the debate described in the introduction. Indeed, this calculation supports the first claim made in \cite{gralla}: that, at least in our simplified model, the astrophysical parameter uncertainties dominate estimates of the black hole shadow size, making a departure from GR difficult to distinguish from inaccurate emission modeling. This calculation argues against the second claim in \cite{gralla}, however: that EHT estimates of the critical impact parameter need to be calibrated with modified-gravity relativistic MHD simulations. In our simplified model corrections to our pseudo-observable $\vartheta$ due to modifications of the gravity theory are subdominant relative to corrections to $b_c$. This is consistent with the notion that one may test modified theories against GR using GRMHD-calibrated estimates of $b_c$ from EHT data.

\par  

Evidently the spherical accretion model used in this paper is an idealized case, although one that can be treated almost entirely analytically. More work needs to be done to determine whether the findings of this paper hold in more complicated accretion models and for black holes with nonzero spin.  We note that there is earlier relevant work on this question using analytic disk models in the Johannsen metric~\citep{2016PhRvL.116c1101J}.  While this work was in preparation we became aware of work by Younsi et al. and Ozel et al. (private communication) that also addresses some of these issues.   

\section*{Acknowledgments}

We thank Yosuke Mizuno for his many helpful comments while preparing this manuscript. A.B., N.Y., and C.F.G. were supported by NSF grant 20-07936. N.Y. also acknowledges financial support through NASA grants~NNX16AB98G, 80NSSC17M0041, 80NSSC18K1352 and NSF grant PHY-1759615. A.C.-A. acknowledges funding from the Fundaci\'on Universitaria Konrad Lorenz (Project 5INV1) and from Will and Kacie Snellings. Computations were performed on the Illinois Campus Cluster, a computing resource operated by the Illinois Campus Cluster Program (ICCP) in conjunction with the National Center for Supercomputing Applications (NCSA), which is supported by funds from the University of Illinois at Urbana-Champaign.

\begin{appendix}

\section{Evaluating analytic profiles interior to the photon orbit in GR, sGB and RZ} \label{app:interior}

In this section, we evaluate the intensity profile in the region interior to the critical impact parameter, i.e., \eqref{eq:interiorfinal}, in GR, sGB and RZ. Before doing so, for brevity, we define
\begin{equation}
    \psi^2 := 27 - b^2.
\end{equation}
Therefore, \eqref{eq:interiorfinal} evaluates to 
\begin{equation}
    I_{GR}^{int}(\alpha,\psi) = \dfrac{2 j_{0} r_{0}^{\alpha} 3^{7/2-\alpha}}{\left(9 + \sqrt{2} \psi\right) \mathcal{C(\psi)}^{1/2}}, \label{eq:GRINT}
\end{equation}
where
\begin{equation}
    \mathcal{C}(\psi) := 2 \psi^4 + 29 \sqrt{2} \psi^3 + 279 \psi^2 + 648 \sqrt{2} \psi + 1458.
\end{equation}
For sGB, we have, to linear order in $\zeta$,
\begin{equation}
    I_{sGB}^{int}(\alpha, \psi) = I_{GR}^{int}(\alpha, \psi) + \tilde{I}_{sGB}^{int}(\alpha, \psi) \zeta,
\end{equation}
where
\begin{equation}
    \tilde{I}_{sGB}^{int}(\alpha, \psi) := \dfrac{j_{0} r_{0}^{\alpha} \sqrt{\pi} \mathcal{P}(\alpha, \psi)}{3^{9/2+\alpha} 10 \psi (9 + \sqrt{2} \psi)^{2} \mathcal{C}(\psi)^{3/2}}, \label{eq:SGBCorInt}
\end{equation}
and 
\begin{align}
    \mathcal{P}(\alpha, \psi) & := 3115661436 \sqrt{2} \\  \nonumber 
    & + (1309851162 + 680953068 \alpha) \psi \\ \nonumber 
    & + \sqrt{2} (378307260 \alpha  - 530129529)\psi^{2} \\ \nonumber 
    & + (197560458 \alpha  - 630061740) \psi^{3} \\ \nonumber 
    & + \sqrt{2} ( 28022760 \alpha - 123277491)\psi^{4} \\ \nonumber  
    & + (3943944 \alpha - 20790954) \psi^{5} \\ \nonumber 
    & + \sqrt{2} (103788 \alpha - 624964) \psi^{6}.
\end{align}
As for RZ, we have, to first order in $a_1$,
\begin{equation}
I_{RZ}^{int}(\alpha,\psi) = I_{GR}^{int}(\alpha,\psi) + \tilde{I}_{RZ}^{int}(\alpha, \psi) a_1,
\end{equation}
where 
\begin{equation}
    \tilde{I}_{RZ}^{int}(\alpha, \psi) := \dfrac{4 j_{0} r_{0}^{\alpha} \sqrt{\pi} \mathcal{U}(\alpha, \psi)}{3^{1/2+\alpha} \psi \left( 9 + \sqrt{2} \psi\right)^{2}\mathcal{C}(\psi)^{3/2} }, \label{eq:RZCorInt}
\end{equation}
and
\begin{align}
   \mathcal{U}(\alpha, \psi) & := 708588 \sqrt{2} \\ \nonumber 
   & + 13122 (43 + 6\alpha) \psi \\ \nonumber 
   & + 3645 \sqrt{2} (11 + 12 \alpha) \psi^{2} \\ \nonumber 
   & + 162 (141 \alpha - 263) \psi^{3} \\ \nonumber
   & + 135 \sqrt{2} (24 \alpha - 83) \psi^{4} \\ \nonumber
   & + 6 (76\alpha - 341 ) \psi^{5} \\ \nonumber 
   & + 4 \sqrt{2} (3 \alpha - 16) \psi^{6}.
\end{align}

\section{Evaluating analytic profiles exterior to the photon orbit in GR, sGB and RZ} \label{app:intensity}

In this section, we evaluate \eqref{eq:intensityclose} in GR, sGB, and RZ, for easy reference. We begin with the GR profile, 

\begin{align}
    I_{GR}^{ext}(\alpha, r_{min}) & = j_{0} \left( \dfrac{r_{0}}{r_{min}} \right)^{\alpha} \dfrac{\sqrt{2 \pi} (r_{min} - 3)\mathcal{S}(\alpha, r_{min})}{r_{min}^{5}}\nonumber \\
    & \times \mathcal{Y}(\alpha, r_{min})^{3/2}, \label{eq:grprofile}
\end{align}
where
\begin{equation}
    \mathcal{S}(\alpha, r_{min}) := \alpha r_{min}^{2} - 2 (\alpha + 6) r_{min} + 36,
\end{equation}
\begin{equation}
    \mathcal{Y}(\alpha, r_{min}) := \dfrac{r_{min}^{3} (r_{min} - 2)^{2} \mathcal{M}(\alpha, r_{min})}{(r_{min} -3 ) \mathcal{S}(\alpha, r_{min})^{2}},
\end{equation}
and 
\begin{equation}
    \mathcal{M}(\alpha, r_{min}) := \alpha r_{min}^{2} - 2 (\alpha + 4)r_{min} + 24.
\end{equation}

For sGB, we have, to linear order in $\zeta$,
\begin{equation}
    I_{sGB}^{ext}(\alpha, r_{min}) = I_{GR}^{ext}(\alpha, r_{min}) + \tilde{I}_{sGB}^{ext}(\alpha, r_{min}) \zeta,
\end{equation}
where
\begin{align}
    \tilde{I}_{sGB}^{ext}(\alpha, r_{min}) & := j_{0} \left( \dfrac{r_{0}}{r_{min}}\right)^{\alpha} \sqrt{\dfrac{\pi}{2}} \nonumber \\
    & \times \dfrac{\mathcal{Y}(\alpha, r_{min})^{3/2}\mathcal{E}(\alpha, r_{min})}{30 r^{11} (r_{min} - 2) \mathcal{M}(\alpha, r_{min}) } \label{eq:mathcale}
\end{align}
and 
\begin{align}
    \mathcal{E}(\alpha, r_{min}) & := 9538560 - r_{min} (12566016 + 1497600 \alpha) \nonumber \\
    & + r^{2}_{min} (6507648 + 2262144 \alpha + 15360 \alpha^{2}) \nonumber \\
    & - r^{3}_{min} (1308864 + 1289408 \alpha + 18752 \alpha^{2} ) \nonumber \\
    & + r_{min}^{4} (696768 + 319888\alpha + 11168 \alpha^{2}) \nonumber \\
    & - r_{min}^{5} (612000 + 148344 \alpha - 376 \alpha^{2}) \nonumber \\
    & + r_{min}^{6} (208224 + 126944 \alpha - 5056 \alpha^{2}) \nonumber \\
    & - r_{min}^{7} (32160 + 44076 \alpha - 1790 \alpha^{2}) \nonumber \\
    & + r_{min}^{8} (2400 + 6460 \alpha + 280 \alpha^{2}) \nonumber \\
    & - r_{min}^{9} \alpha (480 + 215 \alpha) + 30 \alpha^{2} r_{min}^{10}.  
\end{align}
Note the presence of the $(r_{min} - 2)$ term in \eqref{eq:mathcale} causes no pathology in our intensity profile, since $r_{min} > r_{ph}$, and $r_{ph} > 2$.
\par 
Lastly, for RZ, we have
\begin{equation}
    I_{RZ}^{ext}(\alpha, r_{min}) = I_{GR}^{ext}(\alpha, r_{min}) + \tilde{I}_{RZ}^{ext}(\alpha, r_{min}) a_{1}, 
\end{equation}
where
\begin{align}
    \tilde{I}_{RZ}^{ext}(\alpha, r_{min}) & := 2 \sqrt{2 \pi} j_{0} \left( \dfrac{r_{0}}{r_{min}}\right)^{\alpha} \nonumber \\
    & \times \dfrac{\mathcal{Y}(\alpha, r_{min})^{3/2} \mathcal{Q}(\alpha, r_{min})}{r^{8} \mathcal{M}(\alpha, r_{min})}, \label{eq:rzcorrection}
\end{align}
and
\begin{align}
    \mathcal{Q}(\alpha, r_{min}) & := r_{min}^{5} \alpha (5 \alpha - 24) \nonumber \\
    & + r_{min}^{4} (96 + 212 \alpha - 38 \alpha^{2}) \nonumber \\
    & + r_{min}^{3} (92 \alpha^{2} - 532 \alpha - 864) \nonumber \\
    & + r_{min}^{2} (2016 + 192 \alpha - 72 \alpha^{2}) \nonumber \\
    & + r_{min} (432 \alpha + 864) - 5184.
\end{align}
\end{appendix}

\end{document}